## 湯川胸像除幕式への湯川秀樹先生の高知訪問：生涯の転機

### The Visit of Prof. Hideki Yukawa to Kochi to Attend the Unveiling Ceremony of the Yukawa Bronze Statue: Turning Point in Life

Shigeo　Ohkubo[i]

Research Center for Nuclear Physics, Osaka University, Ibaraki 567-0047, Japan

大久保茂男　大阪大学核物理研究センター

　We have discovered unknown very valuable old photos of Prof. Hideki Yukawa at the unveiling ceremony of his bronze statue at an elementary school in Kochi in 1954, which are publicized here. They were discovered at the house of one of the two school pupils who unveiled the statue. We found that the statue had been in front of the main entrance of the school until 1981. However, the statue was moved to a newly created garden near the entrance gate when the new schoolhouse was rebuilt in 1982. The trees in the garden, which grew tall over the statue in the subsequent decades aided and accelerated the forgetting of its existence. The forgetting of the statue by the local people is related by chance to a historic event in the Marshall Islands. Prof. Yukawa's visit to attend the unveiling ceremony occurred just after news that Japanese fishermen were heavily injured by atomic testing at the Bikini Atoll in the Pacific Ocean. At the news conference when he arrived at Kochi, he was unexpectedly asked about his attitude to the bomb testing by the US. He was considered to be the foremost top scientist in nuclear science in Japan at that time, having been awarded the Nobel Prize for physics in 1949. People in Kochi were very anxious about the test since many fishermen worked near Bikini Atoll. Contrary to the expectation of local people, he refused to comment on the matter at the news conference and in public lectures, which were held several times during his stay in Kochi. After returning to Kyoto, he issued the famous statement titled "Nuclear Era and the Turning Point of Mankind" in an influential national newspaper. This was a time in Prof. Yukawa's life when he had to reflect and consult his conscience, and he chose to take a principled stance that became a turning point in his life. After the visit to Kochi, he devoted himself with his wife to the peace movement against nuclear weapons. Looking back, it can be observed that the visit in 1954 to attend the unveiling ceremony of his bronze statue in Kochi marked the start of that change. The statue can be seen to symbolize not only his Nobel Prize but also the turning point in his life that led him to support the peace movement against nuclear weapons.

## 1　はじめに

　イギリスの物理学者チャドウイック(James Chadwick 1891(明治24)-1974(昭和49))が1932年中性子を発見し[1], それをうけ，ドイツの物理学者ハイゼンベルグ(Werner K. Heisenberg 1901(明治34)-1976 (昭和51))[2]とソ連・ウクライナの物理学者イワネンコ(Dmitri D. Ivanenko 1904(明治37)-1994(平成6))[3] は同年ただちに原子核が陽子と中性子からなるという原子核理論を発表した．それ以来物理学において原子核の構成粒子である陽子や中性子をひきつける力である核力がどう生み出されるかが大問題であった．湯川秀樹先生(1907(明治40)-1981(昭和56))はそのしくみを1935年日本の学会誌に発表した「中間子論」[4]で理論的に解明した, 未知の新粒子である中間子が核力を引き起こすとする理論である．　予言された粒子は宇宙線の中に探索され, 1938 年には湯川粒子, ユーコンなどとよばれ, 湯川は一躍世界的に注目されるようになる．湯川の予言した核力を媒介する中間子はイギリス・ブリストル大の物理学者パウエル(Cecil F. Powell 1903(明治36)-1969(昭和44))らにより 1947 年にボリビア・アンデス山脈チャカルタヤ山（標高 5393 メート

ル, Chacaltaya）に設置された写真乾板に飛来した宇宙線のなかに「パイ中間子」として発見され[5], 中間子理論の正しさが明確になった. 新粒子導入の中間子理論で湯川は素粒子物理学をきりひらき, 1949（昭和 24）年日本人として初めてノーベル物理学賞を受賞し, 敗戦で焦土と化し食糧難に喘ぎ打ちひしがれていた国民に勇気と希望と自信をあたえた. 原子核を発見したアーネスト・ラザフォード（Ernest Rutherford 1871（明治 4）-1937（昭和 12））が出身国ニュージーランドの人々の誇りであり, 原子構造を解明したニールス・ボーア（Niels Bohr1 1885（明治 18）-1962（昭和 37））がデンマーク人の誇りであったように, 湯川秀樹とその世界的偉業は日本人の誇りであった. 湯川は生涯にわたって国民に敬愛されつづけた.

湯川の偉業をたたえる銅像は京都大学基礎物理学研究所, 湯川記念館まえの庭に没後5年の1986年京都大学によって公式に設立された. 大学当局が銅像建立に前向きであったのかなかったのか, ノーベル賞を記念して設立された基礎物理学研究所の湯川記念館がいずれ湯川神社になるのではという神社論議[6]が影響したのか, しなかったのかはわからないが, 湯川の銅像が生前京都大学に設立されることはなかった.

ところが湯川の胸像は湯川の生前に建立されていた. 湯川像は1954年高知県の寒村にある小学校に住民運動で建立され, しかもアメリカから帰国間もない湯川先生と夫人が除幕式に出席した. 筆者はこのことを発掘した先稿[7]で湯川胸像の設立経緯, なぜ湯川先生が除幕式に出席したかを詳細に記した.

その後, その湯川胸像建立の除幕式に出席し湯川の話を聞き除幕を行った当時の小学6年生の存在が判明し, また除幕する湯川胸像をご覧になっておられる湯川先生と澄子夫人の鮮明な写真がその小学生に配られていて保存されていたことがわかった[8]. これらは世にはまったく知られてない. 湯川先生は高知の胸像のことを存命中だれにも語らず, どこにも記さなかった. 住民運動によって建立された湯川胸像・設立趣旨は地元でもひとりびとの記憶から時の経過とともに薄れていく. 胸像除幕式での湯川夫妻, 除幕式で湯川の話を聞いた児童・生徒・教職員, 湯川胸像の変遷について新たに見つかった写真をもとに記述. 設立運動の中心となった川村晴吉の功績にも触れる. 湯川博士にとって湯川胸像建立除幕式出席の高知訪問はノーベル賞受賞者としての祝賀歓迎・賛辞一色でもなかったこともわかる. 湯川の高知訪問には訪問1週間前に突如としてあらわれる予期されない水爆の時代のうねりが大きくかかわってくる. 湯川胸像除幕の高知訪問は湯川のその後の物理学者としての人生の転換点となる契機をも孕んでいた. 水爆と原子力の歴史の大きなうねりが重なる.

## 2　高知県夜須小学校湯川胸像の除幕小学生みつかる

京都大学の湯川像建立より 32 年も前に, 高知県の一寒村の小学校に住民運動で設立された湯川胸像は当時の高知県の住民, 地元関係者には当然知られたことではあった. しかし高知県外ではまったく知られないまま 65 年の歳月がながれた. （湯川の 1954 年の高知訪問は湯川著作集の年譜[9]にも載ってない）. 建立後経年とともに残念ながらその地元でも建立関係者は高齢化しまた世を去り次世代に語り継がれることなく記憶はうすれていった. 65 年の歳月を経ると湯川胸像建立の趣旨・経緯だけでなくその存在すら忘れられていった. 稿[7]で PTA を中心に地元住民の自発的運動で建立された日本で初めての湯川博士胸像が高知県夜須町の夜須小学校に存在することが紹介され, その湯川胸像は地元だけでなく国内外に知られることとなった. 敗戦後高知市で生まれた筆者も高知県に湯川胸像が住民運動ではじめて建立され, しかも日本が主権を回復し独立する前の占領時代に渡米し, 5 年間の米国生活から帰国後間もない湯川秀樹博士と澄子夫人（1910（明治 43）-2006（平成 18））[10]が出席して除幕式が行われていた事実を知って驚愕した. 戦後,「日本で天皇に次いで有名」[12]とされた湯川秀樹博士の胸像が 60 年以上も世に知られないままであったことも驚いた.

米国滞在から帰国して間もない湯川先生にとって, 1954（昭和 29）年 3 月の高知訪問は帰国後初めての地方小学校の訪問であった. 日本で初めてのノーベル物理学賞受賞者ということで, 高知では訪問の数か月前から新聞報道され大変な歓迎をもたれ, 到着時も大きな歓迎をうけた. 日本ではじめての湯川胸像の除幕式に湯川先生は澄子夫人とともに出席し, 地元新聞で大きく報道された. その除幕式にのぞむ湯川先生の新聞写真を前稿[7]で紹介した.

拙稿[7]を読み関心をもった新聞記者が高知県香南市夜須町を訪ね, もちまえの記者魂で, 湯川像除幕式で除幕をした当時の夜須小学校の小学生を探しあてた. 若い記者にとって湯川秀樹は学校の国語や社会などの授業で聞くような歴史上の人物であり, 2019年8月私はその記者の求めに応じて新聞社の社屋を訪問し, 記者は除幕式時の湯川秀樹夫妻の写った写真がその小学生に配られた鮮明なまま残されていることをつきとめたと, その写真を見せてくれた（図 1, 図 2, 図 4）. うら若い記者の行動力・探索力に感心するとともに, わたされた名刺に記者が湯川先生と同姓であることに, かつて京都・知恩院にある湯川先生の墓前を訪ねたとき湯川家

頌徳碑に湯川家が和歌山の「湯川城主の末裔」としるされていたことを思い出し，不思議な因縁を感じた．平成生まれの者にとっては第2次世界大戦と敗戦後の70年もの昔の出来事は戦後生れの者が明治時代の日露戦争や乃木希典将軍・東郷平八郎大将を直接知らないのと同様の感覚の遠い昔の歴史上のことであろうか．湯川先生に直接教えをうけたひとはもはや世に少なくなってしまっている．私は夜須町の湯川胸像について拙稿[7]を書くにいたった動機・経緯を話した：筆者は郷里に建立された湯川胸像の存在も知らず当時の若者がそうであったように[13]湯川先生にあこがれ京都大学に進み，湯川先生の教えをうけ，理論物理学・原子核理論の研究者の道に進んだこと，退官後の2013年に文献[14]で記したように湯川先生の四国訪問の事実を知り，湯川先生が高知を訪問したことがあるだろうかに関心を持ったことなど，湯川先生とのかかわり，そして高知の湯川胸像が再発見されたことの意味などについて説明した．湯川先生についての初めて聞く話に同姓の記者は関心をもってくれ耳をかたむけ記録にとどめていた．

2019年春の拙稿[7]の執筆終了と出版後わたし自身はなにか責務を果たし終えたような思いで，湯川胸像のことはすでにあたまから消えつつあった．専門である理論物理学の原子核理論の研究，南部・ゴールドストーンボソンを厳密に取り扱う場の量子論的アプローチでの原子核のアルファ粒子(クラスター)構造の超流動，超固体性についての研究とこれまで続けてきた原子核虹の視点からの中重領域の原子核であるチタン$^{48}$Tiのアルファ粒子(クラスター)構造の研究[15]に没頭していた．2020年初めには日本国内で疫病新型コロナウイルスCovid-19による騒動が勃発[16]，世界が震撼するパンデミックが始まった．疫病の流行は過去の世紀にもたびたびあったようだ．かつてアイザック・ニュートン(Issac Newton 1643(寛永20)–1727 (享保12))はケンブリッジ大学の学生時代に疫病の流行をさけて郷里Woolsthropeに帰り，万有引力の法則に想到したと伝えられ，新進物理学者の第1線にあったハイゼンベルグは1925年疫病の流行をさけHeligorannd島に滞在し，ニュートンの古典力学を根本的に克服する量子力学の新しい着想を得た[17]．

## 3 夜須小学校湯川胸像除幕式での湯川先生夫妻

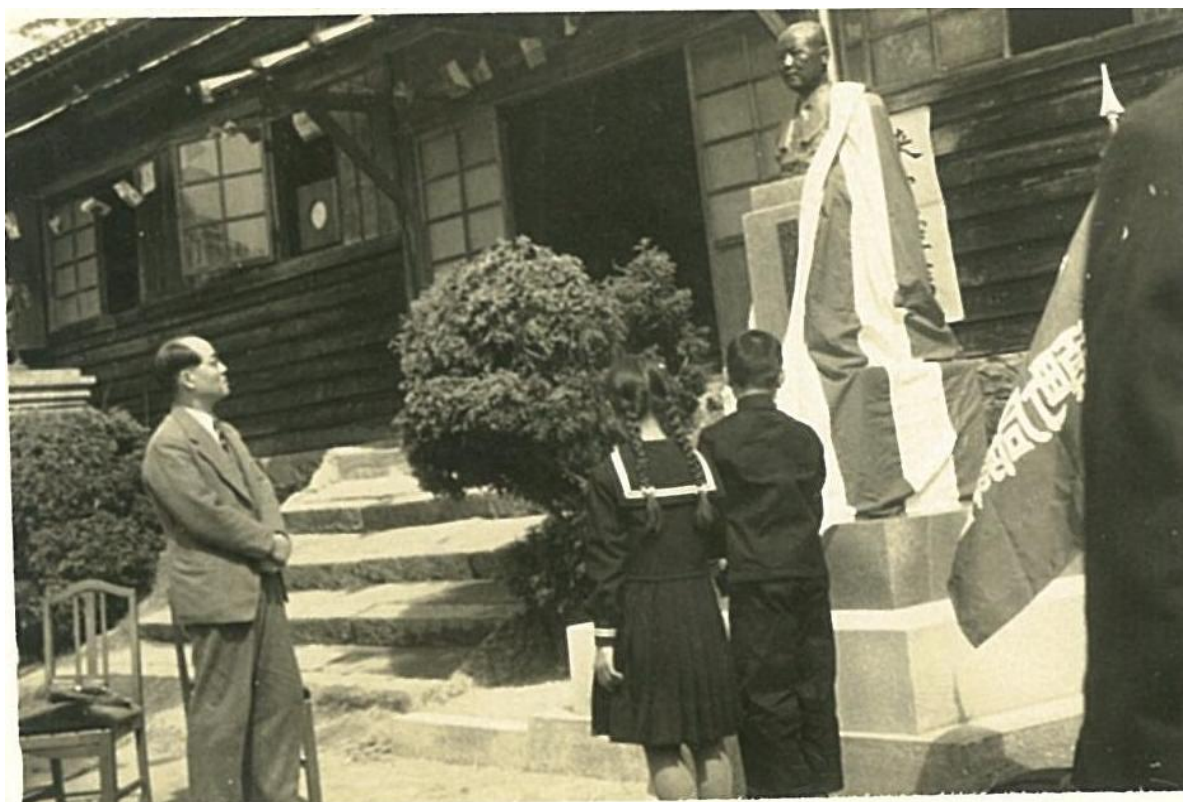

**図 1 湯川胸像(浜口青果(１８９５－１９７９)作)の除幕を見つめる湯川秀樹博士**．除幕するのは小学 6 年の春樹英子(左)と高橋南海男(右)．高知県夜須小学校．1954(昭和 29)年 3 月 22 日．

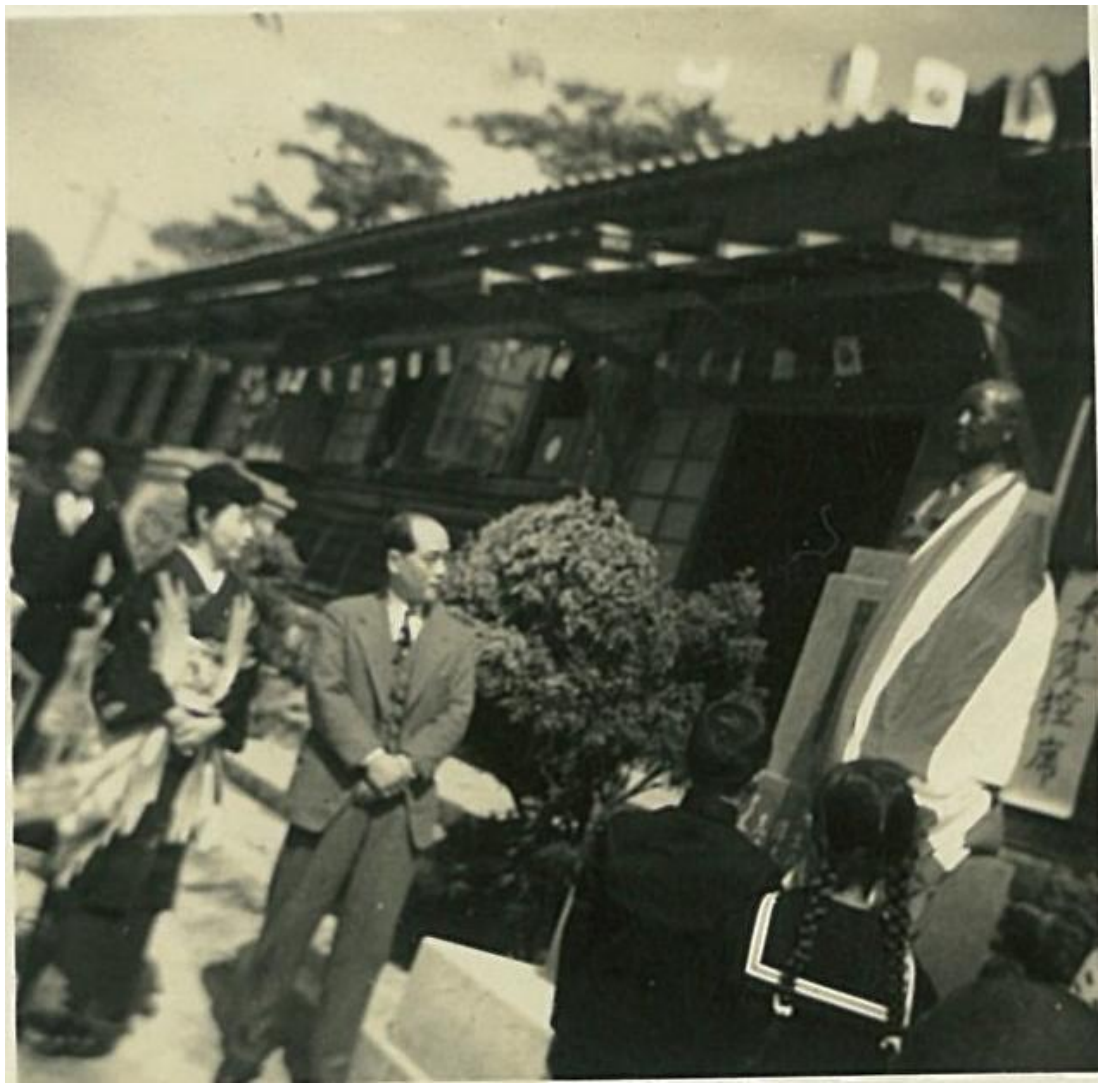

**図 2　除幕の湯川胸像を見つめる湯川秀樹博士と澄子夫人.** 除幕した小学 6 年の高橋南海男(左)と春樹英子(右).
高知県夜須小学校. 1954(昭和29)年 3 月 22 日.

　2021年になって，世界でも日本でもパンデミックが続きひとびとが自粛引きこもり生活を強いられ，筆者も理論物理・原子核理論研究に没入している中[18]，新聞記者がつきとめた湯川胸像除幕式に出席したという当時の夜須小学校6年生から突然電話がかかってきた．電話の主は清藤禮次郎. 私の書いた湯川胸像についてぜひとも説明したい，1954年3月夜須小学校卒業で湯川先生のお話を直接聞いたという. 2日後(2月8日)には資料を持って寒さの厳しいなか遠路はるばる訪ねてきた．さらに，約10日後(2月19日)には清藤禮次郎がさらに詳しい説明をしたいと除幕をおこなった児童で新聞写真に載った浜田(旧姓春樹)英子を伴いアルバムなどたくさんの資料をもって再び訪ねてきた．保存されていた除幕式での湯川先生の写真の現物(図1-図4)や卒業アルバムなど資料をたくさん見せてくれ，当時の模様を思い出し語ってくれた．夜須小学校の湯川胸像のことが忘れ去られることなく伝えられ残ってほしいとの思いがつたわってきた．2019年5月に先の原稿[7]を執筆・出版公表された時にはまったく予想しないことであった．世のめぐりあわせと因縁の不思議に驚くとともに，教えをうけた湯川先生の墓前に報告したいような思いで，また探し当てた新聞記者・当時の卒業生，湯川胸像を建立した人びとの思いにおかれる気持ちで本稿をしるす思いにいたった次第である．

　湯川胸像を除幕したのは1954年3月卒業，当時6年生であった6年C組の春樹英子と6年A組の高橋南海男．高橋

南海男は2020年に泉下の人となったときき，話を聞く機会が失われたのはまことに残念である．夜須小学校での湯川胸像除幕式は1954(昭和29)年3月22日午前に夜須町夜須小学校校庭で行われ，今回見つかったのは6年C組の浜田英子にクラス担任小松明先生から配られた写真4葉である．当時は写真をたくさん印画紙に焼き増しすることがかんたんでなく，除幕式をした児童にのみ学校から配られたという．

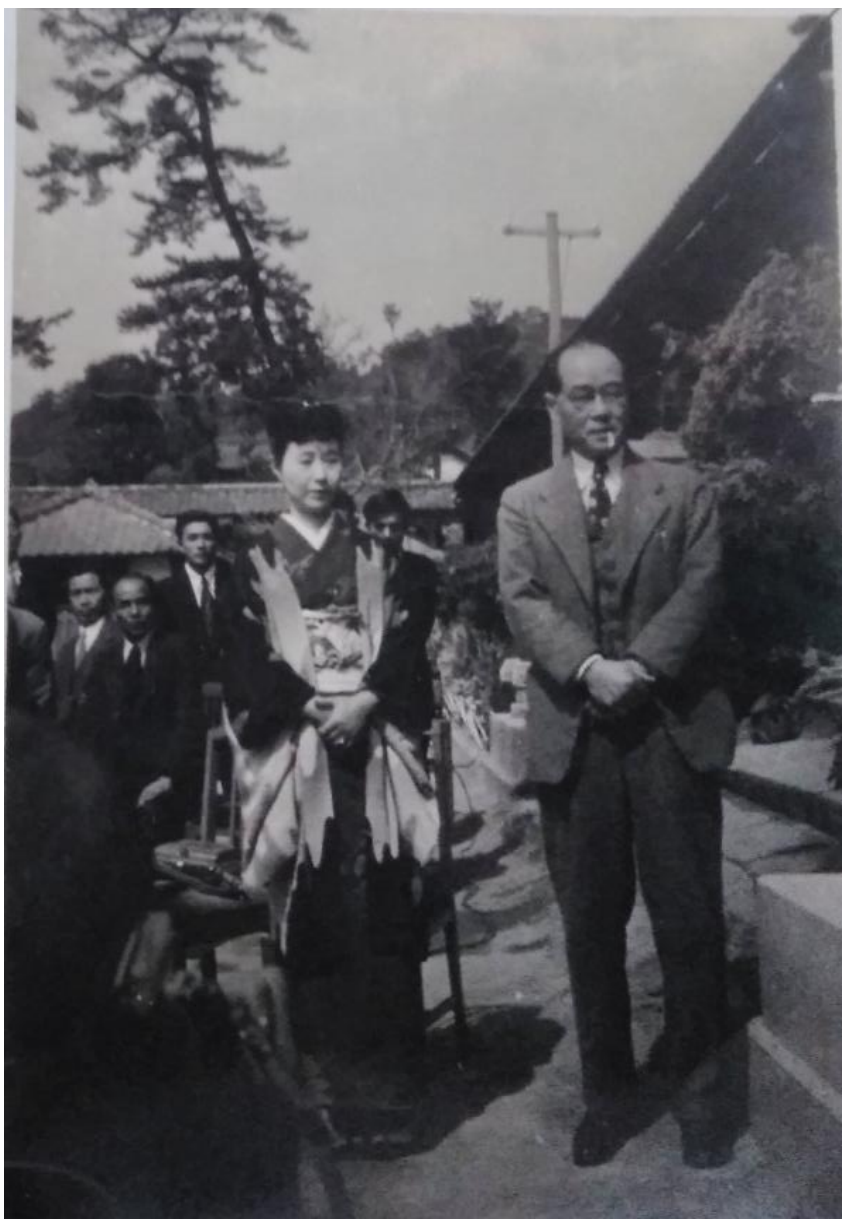

**図3　湯川胸像除幕式の湯川秀樹博士と澄子夫人．後ろの建物は夜須小学校校舎．** 1954(昭和29)年3月22日.

除幕式の当日は春分の日で祝日にもかかわらず夜須小学，夜須中学の児童・生徒全員が朝から登校し，春樹英子は除幕のやりかた・手順について先生から説明をうけた．「えらい先生」だと担任の小松明先生から聞かされた春樹は除幕で緊張した．除幕の際も後も湯川先生夫妻と直接お話しすることはなかった．だが，春樹には目の前に見る若々しくはつらつとしたアメリカ帰りの湯川夫人の上品な女性という印象はつよくのこった．67年後のいまでもよく覚えている[19]．夜須小学校と夜須中学校の児童・生徒900人は全員校庭に整列し，除幕を見

守った．除幕式をとりおこなった近藤亘校長は1949年4月に夜須小学校に赴任している．図 1は卒業する6年生春樹英子と高橋南海男によって除幕される胸像を見上げる湯川秀樹先生．木造校舎には万国旗が飾られている．図 2は除幕された胸像を見つめる湯川秀樹先生と湯川澄子夫人．図 3は胸像横に立つ湯川先生ご夫妻で後方には来賓など関係者がみえる．図 4は除幕式での壇のまえのご夫妻，後方には二宮金次郎の像がみえる．湯川先生はこのあと壇上でお礼のあいさつを行っている．「きょうはこんなにりっぱな胸像をたてていただいて感慨無量です，私はあたり前の人間にすぎないが根気よく勉強をつづけただけのことです，この胸像がみなさんの役にたてば大変光栄です」[20]．

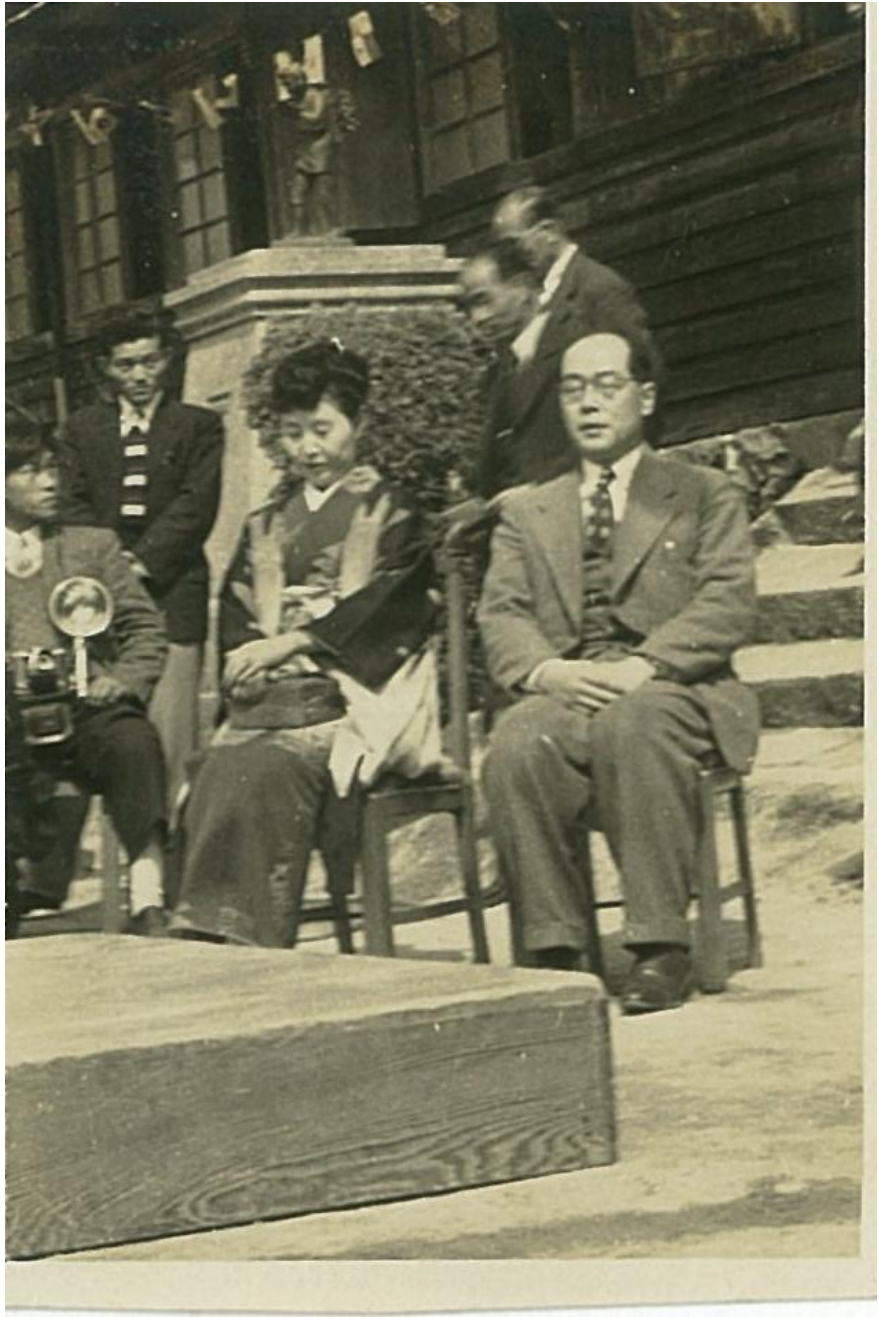

**図4　湯川胸像除幕式の湯川秀樹博士と澄子夫人．** 前の壇上で胸像建立に対するお礼のあいさつを行った．高知県夜須小学校．1954 (昭和29)年3月22日．

　図5は湯川胸像除幕式に出席し湯川先生の話を聞いた6年生の全員の卒業写真である．除幕式が行われた校舎正面玄関前で写されている．湯川胸像建立は夜須小学校の子どもたちへの科学教育の振興のためにこの6年生の卒業にあわせて小学校PTAにより制作が企画された．除幕式の日も卒業に合わせて日程が決められたのである．図6は卒業式当日の教職員の写真である．近藤校長とならんで湯川胸像設立運動の中心となってきた夜須小学校PTA会長川村晴吉翁が写っている．図7は児童と教職員の写真で，賞状をもつ6年生春樹英子と春樹に除幕式の手順を指導したC組担任の小松明らが写っている．（この写真は卒業式とは別の日に高知県合唱コ

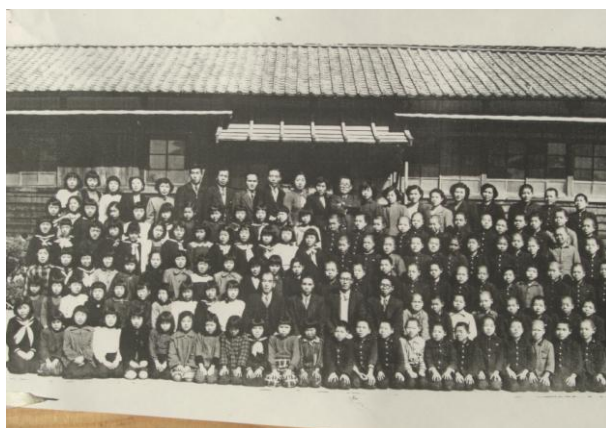

**図5**　1954(昭和29)年3月湯川胸像除幕式で湯川秀樹の話を聞いた高知県夜須小学校6年生123人の学校正面玄関での卒業写真．湯川胸像は玄関右．前2列左から11番目が近藤亘校長．除幕したのはA組の高橋南海男（前4列目右の端）とC組の春樹英子（前3列目左から6番目）．B組の清藤禮次郎は前5列目右から5番目．

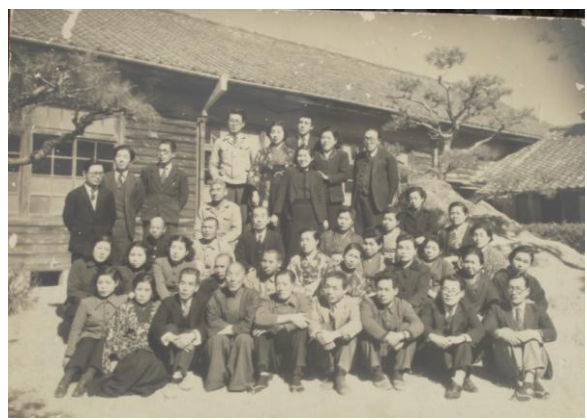

**図6**　1954(昭和29)年3月湯川胸像除幕式で湯川秀樹の話を聞いた高知県夜須小学校教職員．6年生の卒業式のさいに南校舎と北校舎との間の中庭で撮影．前列左より3人目近藤亘校長，4人目夜須小学校PTA会長川村晴吉．

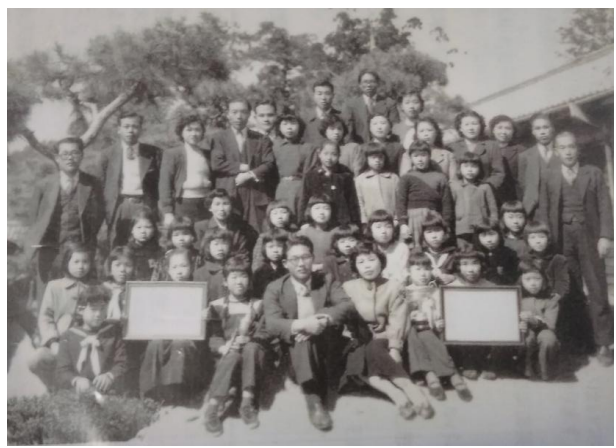

**図7**　高知県夜須小学校の生徒と教職員の写真．春樹英子(前列左より3人目，賞状をもつ)に湯川胸像の除幕の手順などを指導した6年C組の担任小松明先生(前列中央，その右は石本先生)．後列左より4人目に石田教頭，2列目右端は近藤亘校長．

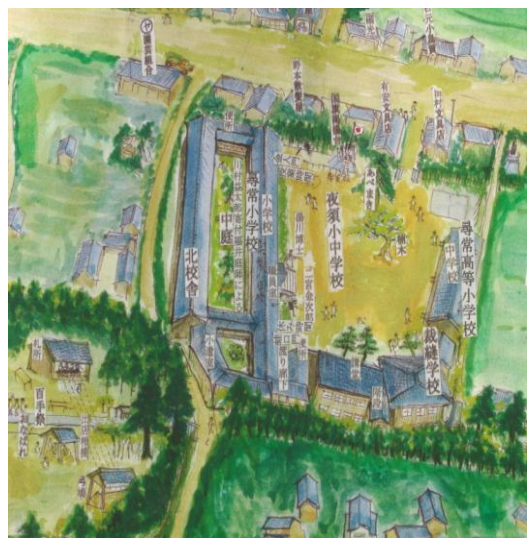

**図8**　1954(昭和29)年頃までの高知県夜須小学校の校舎配置図絵(清藤禮次郎作)．湯川博士，二宮金次郎と書かれているのが見える．

ンクールでの優勝のさいに写されている．）

　湯川胸像の存在は清藤禮次郎が書いた『私の昭和』にある昭和 29 年頃までの夜須小学校の校舎配置図を描いた絵画（図 8）にも見ることができる．尋常小学校の前に「湯川博士」「二宮金次郎」と書かれているのが見える．湯川の訪問は夜須小学校の子どもの進路にも影響を与えたようだ．湯川の話を聞いた児童のなかには湯川にあこがれた京都大学理学部に進んだものもいる[21]．

## 4　湯川胸像建立を計画した川村晴吉翁と土佐の気風

　湯川胸像設立の発起人で中心になったのはPTA会長をながくつとめた川村晴吉（1876（明治9）−1974（昭和49））である．1876（明治9）年，現夜須町坪井に生まれ1974（昭和49）年2月1日満97歳で死去．夜須小学校への長年の貢献で，夜須小学校図書室には「川村記念文庫」がもうけられ，川村晴吉の胸像がある．夜須町史には「川村記念文庫」についてつぎのように記され貢献が讃えられている[22]．「川村記念文庫とは，川村晴吉翁の寄贈により

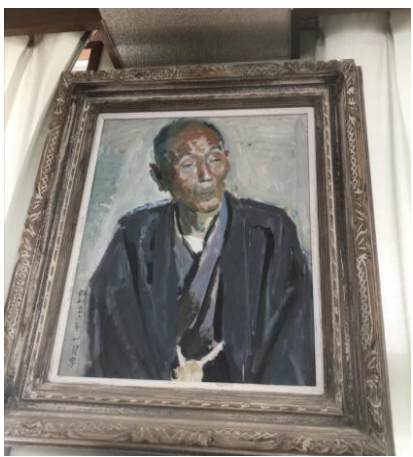

図 9　湯川胸像建立提唱の高知県夜須小学校 PTA 会長川村晴吉肖像画．夜須小学校図書室所蔵（上島一司（1920−1994）作）．

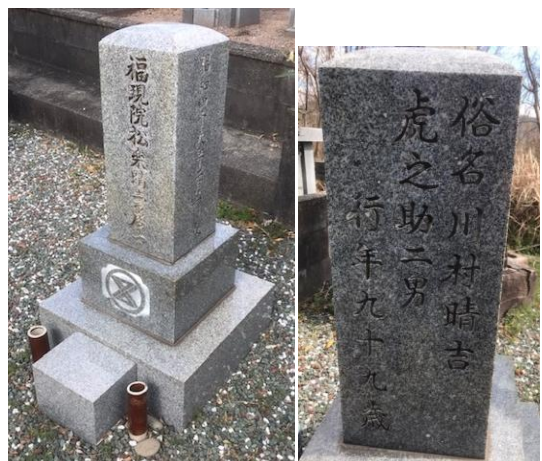

図 10　夜須小学校湯川胸像建立提唱の川村晴吉墓碑．虎之助次男行年 99 歳とある．

作られたもので，昭和二九年九月夜須小学校図書館内に併設された．川村晴吉翁は昭和五年六月二〇日夜須小学校教育後援会長となって以来，二七年一一月夜須町教育委員会の委員になるまでの二二年間，戦前はもちろん戦中・戦後の激動の中に有って後援会長・PTA会長として，学校教育への協力，児童の福祉増進，父母の啓蒙などに尽くし，その後も町教育委員として十年間その職にあり，あわせると三〇有余年間夜須町教育に尽くした功績はまことに大きい．町はこれにこたえる（ママ筆者）に小学校玄関脇に上島画伯の手になる肖像画をかかげるとともに盛大な謝恩会を開いた．また三八年一一月九日には永く翁の栄誉をたたえるために胸像を建立した．」肖像画は図 9である．川村晴吉翁の胸像は現在夜須小学校図書室にある．名前については，新聞などは「川村春吉」とされているが[7]，ただしくは夜須町の墓碑（図 10）にあるように川村晴吉である．川村晴吉は結婚していて子供がいた．川村晴吉の子どもたちの教育，学校での科学振興へのつよい情熱と献身とリーダーシップによって夜須小学校の湯川胸像建立が実現した．

　その実現には川村の「湯川胸像」建立の提唱に賛同しいっしょに進めた小学校PTA・住民の運動のはたした役割も大きい．住民運動で「湯川胸像」が設立されるのには時代をきり開いてきた先人をたたえる土佐の進取の気風があるように思われる．足摺岬の四国88カ所38札所・金剛福寺（土佐清水市）の近くには日本人として最初にアメリカにわたり教育をうけ，帰国後は河田小龍（1824（文政7）−1898（明治31））に海外の時代状況をしらせ，坂本竜馬（1836（天保6）−1867（慶応3））に世界へ飛躍する雄志をいだかせた土佐清水市出身のジョン・万次郎（中濱万次郎）（1827（文政10）−1898（明治31））の銅像がある．万次郎は後に開成学校，東京大学の教授となる．高知市・桂浜に太平洋を望む坂本竜馬の銅像も住民運動と募金で建てられた．室戸岬から太平洋をのぞむ坂本竜馬

の同志・中岡慎太郎(1838(天保9)-1867(慶応3))，須崎市・横浪三里に太平洋をのぞむ土佐勤王党の武市半平太(1829(文政12)-1867(慶応元年))の銅像もそうである．土佐の思想家・植木枝盛(1857(安政4)-1892(明治25))の「自由は土佐の山間から」という気風が住民運動の根底に流れているようだ．教育にかかわる偉人としては足摺岬近くの宿毛市には「学問の独立」を唱え大隈重信（1838(天保9)-1922(大正11))とともに早稲田大学を創立した小野梓(1852(嘉永5)-1886(明治19))の銅像が早稲田大学の寄贈により生家跡の小野梓記念公園に，また「小野梓君碑」が清宝寺に建てられている．最近になっては2019年に高知市に物理学者で日本におけるX線物理学の魁である[23]寺田寅彦(1878(明治11)-1935(昭和10))の銅像が住民運動により建立されている．物理学者の銅像としては湯川胸像につぐ高知県内2番目の住民運動による銅像である．寺田寅彦はX線が波動で回折・反射することを示し[23,24]，ノーベル賞級の仕事を行い，数か月の差でノーベル賞を逸したとはいえ，日本のX線結晶解析物理学の先駆的な科学者[23]として住民によっても顕彰されるのはもっともなことである．寺田につづく菊池正士の電子線回折の研究もノーベル賞級と言われている．菊池正士(1902(明治35)-1974(昭和49))は寺田の教え子で[26]，寺田は菊池を通じ湯川にも間接的に影響を与えているかもしれない[27-28]．川村晴吉は寺田寅彦より2歳年長，同世代だが寺田より40年近く長命し数え年齢99歳で歿，湯川胸像建立時には78歳．寺田は文豪・夏目漱石(1867(慶応3)-1916(大正5))の小説『吾輩は猫である』の水島寒月や『三四郎』(1909年出版)の野々宮宗八の科学者のモデルとしても登場し，また自身多くの随筆や名言で文学にも足跡を残しているので，川村晴吉も同郷の寺田から科学や科学振興に興味をそそられる刺激や影響をうけていたかもしれない．これらの運動や土佐の進取の気風をみると湯川のノーベル賞受賞を顕彰し銅像を建てようという運動が戦後の貧しく困難な中で高知に起こったというのも決して偶然とは言えないように思える．

## 5 夜須小学校湯川胸像の位置の変遷と忘却

　高知の湯川胸像は建立後も他県や国外に知られることなく，当の学校でも語り継がれることなく，教職員らにもその設立経緯は忘れられていく．いつごろまで児童や職員に記憶されていたであろうか．胸像建立後の学校校舎改築の変遷をみると，胸像の場所が動いていたことがわかる．湯川胸像の除幕式の当時夜須小学校 6 年生であり除幕を行った春樹英子と同学年級友であった清藤禮次郎(図5の1954年3月の卒業写真に写っている)が資料を提供してくださった．湯川胸像が建立されたときの旧木造校舎は1958(昭和33)年に改築され新しい木造校舎になっている．図11，図12は新木造校舎の遠景である．図13は1981年の夜須小学校での卒業写真である．児童らの背景の玄関右には湯川胸像がある．湯川胸像は木の陰に隠れることもなく，登下校する児童や教職員，学校を訪れる町民のだれもの目に映っていたことは明らかである．この校舎の時代には卒業写真は湯川胸像を背景に撮られている．

　1982(昭和57)年には新木造校舎も現在の鉄筋校舎に改築される．これにより湯川胸像は玄関右側前にあら

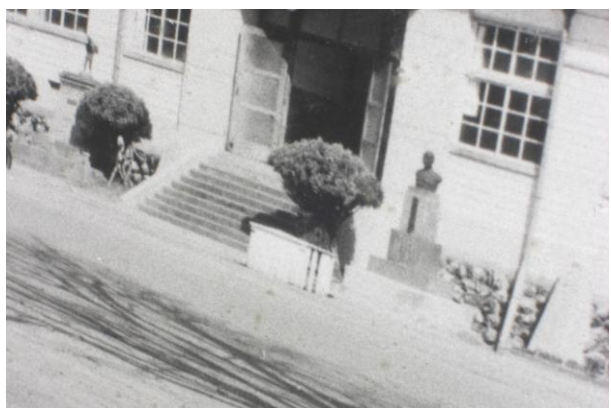

図11　1959(昭和34)年4月撮影の高知県夜須小学校の校舎と湯川胸像(玄関右側)．玄関左側に見えるのは二宮金次郎の銅像．

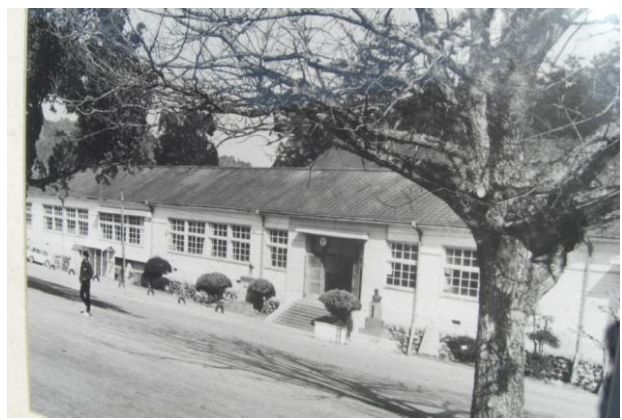

図12　1959(昭和34)年4月撮影の高知県夜須小学校の校舎全景と湯川胸像(玄関右側)．玄関左側には二宮金次郎の銅像．

たに築造の庭である「改築記念園」に移される．二宮金次郎の銅像は玄関左側であり，両者の位置関係は変わっていない．鉄筋校舎建設当時は湯川胸像も児童の登下校時に目に入る位置にあったが，庭の松の木が成長するとその陰に入るようになっていく．特に意識して見ない限り登下校時に湯川胸像が目に入ることはなく，湯川胸像は徐々に児童たちの日常の視界から遠ざかっていった．

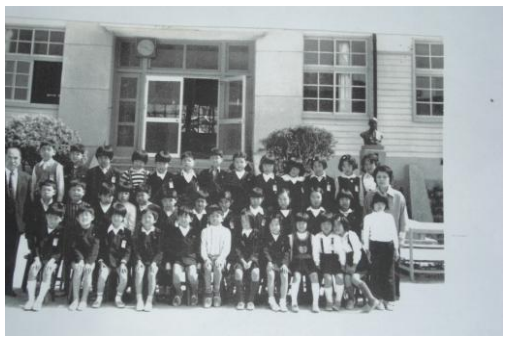

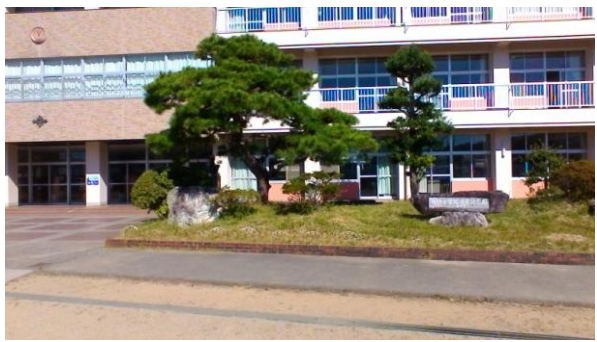

**図13　1981（昭和56）年3月の高知県夜須小学校6年生卒業写真**．校舎は昭和33年に改築，湯川胸像は正面玄関右側．

**図14　1982（昭和57）年改築の高知県夜須小学校鉄筋コンクリート校舎**．改築記念園に湯川胸像がある．2018年11月4日筆者撮影．

　図14は2018年に撮影された鉄筋コンクリート校舎である．二宮金次郎像は玄関左にあり視界に入るが，湯川胸像は庭園の木陰にひっそりとたたずみ人目につきにくい．図13のごとく，校舎正面玄関前で卒業写真を撮ったとしても湯川胸像は背景には写らない．1974（昭和49）年には湯川胸像設立の中心的人物であった川村晴吉が世を去る．川村晴吉翁の逝去と設立運動関係者世代の高齢化，経年とともに湯川胸像は語り継がれることなく学校関係者，地域の人びとの記憶からしだいに薄れていった．また，湯川胸像は1954年建立年以降，筆者により再発見されるまで新聞・メディアなどで一度も報道されることなく，人びとの記憶から消えていった．筆者が2018年，夜須小学校の湯川胸像を確認のため見に行った際，湯川胸像が夜須小学校の児童・住民に知られてなかったことは前の拙稿[7]でふれた．

　この徐々なる記憶の喪失はたんなる経年性によるのか．湯川胸像が人びとの記憶から消え，また湯川博士が語らなかった背景にはもっと大きな歴史のうねり，歴史の転換があるようだ．

## 6　高知到着と湯川の記者会見：湯川胸像と水爆

　湯川先生の生涯のなかでも胸像が建立された1954年3月は「人生観の大転換」のときでもあった．1954年3月が湯川の人生観の転換であったことはその年の講演「私の人生観の変遷」[29]でも語られている．

　そもそも湯川先生は京都から遠隔の地である高知をなぜ訪問したのか[30]．当時は国鉄の列車でも片道1日がかりの長旅である．湯川の高知訪問の経緯を子細に見ると，湯川胸像建立除幕式出席の当初の目的が水爆でややかすんだことが見えてくる．湯川にとっては思いもよらぬ歴史のめぐりあわせである．歴史の波に呑み込まれることの2回目の体験である．1回目の体験は戦争前，大阪帝大助教授のときであり，拙稿[14]で述べた．教育熱心な徳島の中学校長が1938年10月9日講演依頼のため無名の湯川助教授を訪ねるが，湯川は躊躇する．ところが校長が帰ったその日の夕刊にたまたま湯川のことが新聞に大々的に報じられ湯川は一躍時の人として世に躍り出る．湯川はこう書いてい

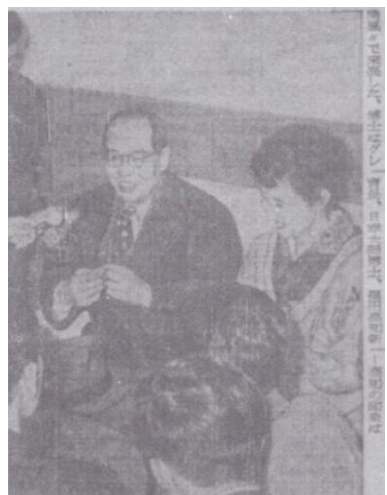

**図15　記者会見する湯川秀樹夫妻**．1954年3月21日国鉄高知駅．3月22日高知新聞[32]．

る．「ところが世の中には不思議なことが起こるものでして，ちょうどその日の夕刊に私の中間子論の研究を，世界的な研究だと，各社の新聞が一斉に，非常に大きく書いたんです．」[31]自信を深めた校長は翌日湯川に一層熱心に講演依頼を行い，湯川はけっきょく引き受け「はじめて一般向け講演」[14]のため徳島を訪れることになる．この講演はその後の生涯を通じての湯川の旺盛な一般向け講演活動の原点となる．偶然のめぐりあわせが第 1 回目が 1938 年であったとすれば，その第 2 回目は湯川が胸像除幕式出席のため高知を訪れた 1954 年 3 月であったであろう．

　湯川高知訪問を詳しく見てみる．湯川は 1954 年 3 月 21 日夕刻列車で高知駅に到着し，高知市長をはじめとする 2000 人を超える大勢の高知市民の歓迎をうけた．湯川にとっては生まれて初めての高知訪問である．翌日 3 月 22 日の高知新聞朝刊[32]の伝えるところによると「一九四九年度ノーベル物理学賞に輝く湯川秀樹夫妻は高知新聞社主催の講演会ならびに夜須小学校同博士胸像除幕式出席のため二十一日午後七時十五分高知駅着"南風"で来高した．博士はグレイの背広に水玉模様のネクタイ，柔かい黒のオーバーを着込み顔色もよく，若々しい澄子夫人はコン地の花模様の和服姿出迎えの氏原市長，高橋県教育長，伊藤市教育長，日赤大岡博士，福田高知新聞社社長，堅田同編集局長，教育関係者，婦人団体，学生ら千名にニコやかに挨拶した．」とある．

　湯川は国鉄高知駅長室で記者団との一問一答で湯川胸像について次のように答えている[32]．

「夜須小の胸像について：
博士　私は皆がいうほど偉人でもなければりっぱな人間でもない．二宮尊徳像にかわって私の胸像建立はどうかと思うが日本では最初のことだし出席することにした．」ついで澄子夫人は科学者の妻として感想を次のとおり語っている．「研究に熱中しすぎて家の中が暗くなり易いのでつとめて明るいふんい気を作るように努めています．主人は胃腸が弱いので食物はとくに注意し，睡眠不足にならぬよう留意しています．中間子の存在も寝床で発見したくらいなので頭をつかれさせぬよう心がけています．科学者の妻といっても主人に来た手紙を整理するくらいことで一般とそんなに変りはありません．」[32] この高知駅での湯川胸像についての記者団とのやりとり，また澄子夫人の会見内容は和やかなものであり，湯川には予想されたことであったろう．

　湯川の高知訪問が持ち上がったのは半年前である．当時の状況をしらべると，湯川胸像の制作は湯川訪問とは別個のこととして湯川の訪問の前年 1953 年 10 月から始まっている[20]．湯川の高知訪問が決まるのは銅像制作開始の後である．1954 年 1 月 13 日の高知新聞[33]は「湯川博士 3 月中に来高」と報じ，1 月になって高知訪問が決まったことを伝えている．2 月 6 日の高知新聞[34]は「湯川博士　来高日程決まる」と報じ，2 月 19 日の高知新聞[35]は「湯川博士の来高本決まり」と報じている．

　湯川は高知を訪問することについて前述のように除幕式との関連では「日本では最初のことだし出席することにした」と新聞にのっているが[32]，敗戦後の混乱と全土が焦土と化し国民全体がまずしい戦後復興のなかで，地方の寒村の住民が厳しい生活のなかで寄付金を出し合い自主的に「胸像」を建設するということに胸を打たれたのではないだろうか．当時のきびしい戦後復興のなかでは住民が寄付をだすということは食費代をけずってでも胸像を建てるということである．除幕式のあいさつで「この胸像がみなさんの役にたてば大変光栄です」と述べているように，住民への感謝の気持ちを抱いて出席を決意したのではないだろうか．こういう運動は日本国内で高知県夜須町以外では起こってない．

　当時の時代背景をみると，アメリカでの 5 年間の生活から日本に帰った湯川が，戦勝国での近代的機械文明の生活とは対照的に，占領軍からの独立間もない戦後復興の厳しい生活におかれている住民の意気に感激し，京都からはるか遠方の寒村に出向く決意をしたのも理解できるように思われる．湯川は除幕式出席だけでなく，いくつかの講演も依頼されていた．3 月 22 日午前の夜須小学校での湯川胸像除幕式出席のあと，午後には夜須町のとなりの赤岡町にある城山高校で講演，夕方には高知市内での一般向け講演，翌 23 日には高知市での午前の児童への講話，午後の学校関係者への学術講演などと，ハードスケジュールの講演が予定されていた．日本で初めてのノーベル賞受賞者，アメリカから帰国後初の地方・遠隔地で講演，高知県へのはじめての訪問，であるから，各方面からの講演の企画はやむをえないことであろう．ひとびとは熱狂的に湯川を出迎えたのである．

　しかし湯川を待ちうけていたのは「湯川胸像」だけではない．湯川の生涯・人生の転機となる大事件が偶然とはいえ湯川を待ち受けていた．しかも人々の当面の関心はそちらの方がはるかに大きかった．湯川を多くの

市民が駅に歓迎に出迎えたのには，日本初の偉大なノーベル賞受賞者を歓迎するというだけなく，遠洋漁業が重要な産業である室戸・土佐清水など漁業のまちの切実な問題もかかわっていた．長旅で到着した湯川をまっていたのは，新聞記者らの厳しい質問であった．高知駅での記者団との一問一答の記者会見で最初にでた質問は，ノーベル賞受賞の感想でも，夜須小学校の湯川胸像についてでもなく，水爆であった．以下に一問一答を高知新聞 [32] より見てみよう．

　　　「一　　高知の印象は
　　博士　本県は初めてです。四国で高知だけ来ていないので今回実現したのはうれしい。
　　　一　ビキニ被爆と政府の原子炉予算について
　　博士　これについてはご承知のように一般に言明したことがなく全く関知しないところで私の研究外だ。この問題は答えられない。
　　　一　原子力の平和利用について
　　博士　原子力の国際管理は必要なことで日本も将来参加の要請があれば参与すべきだ。原子力を平和的に使うべきだということはいまさらいうまでもない。
　　　一　高知での講演内容は
　　博士　原子力についてはふれない、一般的な所見、科学者としての体験につて述べるつもりだ。
　　　一　夜須小の胸像について
　　博士　私は皆がいうほど偉人でもなければりっぱな人間でもない。二宮尊徳像にかわって私の胸像建立はどうかと思うが日本では最初のことだし出席することにした。
　　　一　将来の研究と湯川会館について
　　博士　従来どおりの研究を続ける、また湯川会館は理論物理を研究する同志が利用しうるよう充実さしたい
　　　一　将来渡米の計画は？
　　博士　招へいはあるが行くつもりはなく全部断ってある。」（下線は筆者）

　高知の印象にかんする儀礼的質問のあとの冒頭質問がビキニ水爆であろうとは，湯川の高知訪問が企画されたときには，まったく予想されないことであった．この質問について湯川は「ご承知のように一般に言明したことがなく全く関知しないところで私の研究外だ。この問題は答えられない」と断固拒否している．新聞記者がするどい質問を高知に到着するや否や長い列車での旅で疲れている湯川に投げかけたのには当時の世論があった．

　いうまでもなく 1954 年 3 月 1 日，米国はビキニ環礁で水爆実験を行い，世界を驚愕させた．また，ビキニ環礁水爆実験がおこなわれたと同じ日，1954 年 3 月 1 日に原子炉築造予算が突如国会に提出された．3 月 1 日のアメリカによるビキニ環礁での水爆実験により，マーシャル諸島で操業していた日本の遠洋マグロ漁第五福竜丸(静岡県焼津)の船員 23 名全員が水爆の放射性降下物で被爆し，無線長の久保山愛吉が放射線障害で死亡した．近くには 1000 隻をこえるたくさんの日本の漁船が操業し[36](被災漁船は 856 隻以上)[37]，高知県は遠洋漁業が盛んで多くの漁船が操業していた[36-40]．

　ビキニ水爆実験は 3 月 1 日行われたが，それがすぐに世の中に衝撃を与えたわけではない．1954 年 3 月 1 日からの状況をやや子細に見てみると，最初に新聞報道されるのは 3 月 2 日である．報道には水爆という言葉は出てこない．朝日新聞 3 月 2 日の夕刊[41]は一面ではあるが小さく「新原子爆発実験　マーシャル群島で始まる」と報じているにすぎない．3 月 14 日になって夕刊[42]に「強力水爆を実験　米政府高官ほのめかす」という記事がでている．3 月 16 日に「ハワイへ出発　原子力委員長」という記事が出る[43]．3 月 14 日焼津港に第五福竜丸が帰国しても，すぐ報道され大きな騒ぎが起きたわけでもない．大事件となるのは読売新聞 3 月 16 日朝刊[44]の特ダネ報道である．読売新聞の見出しは「邦人漁夫，ビキニ原爆実験に遭遇　23 名が原子病 1 名は東大で重症と診断」[44]とある．原爆，水爆は原子の力とされ，アメリカでも原子力委員会が管轄し，水爆実験も原子力委員会のもとで予算がつけられ開発実験が行われていた．「原子」と「核」は，はっきりと区別されてつかわれてなく，広義に原子兵器，原子病，原子力とよばれた．3 月 1 日の水爆実験にかんす

る最初の発表もアメリカ原子力委員会のストローズ委員長がおこない「原子力装置」の爆発がマーシャル諸島で行われた，と発表している.

3月16日になって国民ははじめて水爆による日本国民の被爆を知るのである. 湯川は3月16日当日すぐに日記に次のように書いている[45].「3 月16日 火 曇 ＜略＞三月一日ビキニ環礁北東約百マイルの地点で水爆実験による真っ白な灰を被ったマグロ漁船第五福竜丸帰港， 火傷の傷害を受けた乗組員を診断 水爆症と推定」. 湯川は高知へ出発する1週間前に，湯川が5年間滞在し前年の7月までいたアメリカによる水爆実験と国民の被爆を知るのである. 湯川が3月16日についで日記を書くのは3月28日である. 湯川は3月21日-23日高知を訪問・滞在し，この間のことは日記にない. つまり，湯川が湯川胸像の除幕式に出発する1週間前に，日本の世論を大きく変えその後生涯にわたって湯川を巻き込むことになる重大事件が発生したのである. 原子力の専門家，権威とみられている湯川が沈黙をつづけるなかで，記者団が原子力，水爆について第一番目の質問するのは，社会の代表として当然のことであると思われる.アメリカによる水爆実験と第五福竜丸の被害が報じられて初めてマスメディアに対応した3月21日，湯川は質問にたいして断固として一切の答えを拒否し沈黙を守った.

高知駅での大きな歓迎について取材した新聞記者はのちに次のように書いている[46].「"世紀の頭脳"といわれるノーベル賞湯川さんが・・・現れたのだから大向うは沸いた. 高知駅を埋めつくした歓迎の市民は世界の原爆不安を解消してくれる救世主のようなあこがれで湯川さんの顔をあおいだ. ぼくたちも思いははおなじ. 記者団の質問は期せずして原爆問題に集中した. しかし『ノーコメント、私のやっている学問は原爆とは関係ないんだ』と博士は無知な記者どもに顔をしかめた.」

## 7　湯川の高知講演：水爆・原子力と「自己本位」的物理学者の苦悩

アメリカによるビキニ環礁での水爆実験と日本人の被害が3月16日に新聞で報じられてから，湯川がはじめて一般に話すのは，翌日3月22日午前の夜須小学校湯川胸像除幕式でのあいさつである. 先に書いたように，その「あいさつ」では当然，原子力，水爆にはふれていない. 11時半に除幕式を終え，12時半から隣接の赤岡町の県立城山高校でも2000名余の聴衆に一般むけの講演を約40分間行う. 会場に入れなかった聴衆のために屋外で同じ講演をもう一度おこなっている. 講演内容は「私は幼い時から志を立ててこれを実行してきた. 将来もこれで進む覚悟である，あながち偉人になるのがえらいのではない，皆様は志を立てて将来一つの研究に努力を傾け有能な人物になって世界人類のため尽くしてほしい」[47]. 講演では，水爆，ビキニ被爆問題にはまったくふれていない. 午後はかねてからの希望

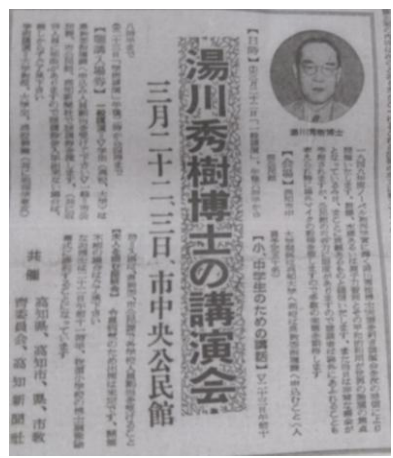

図16　湯川博士の講演案内 1954 年 3 月 13 日（土曜日）.　高知新聞朝刊[48].

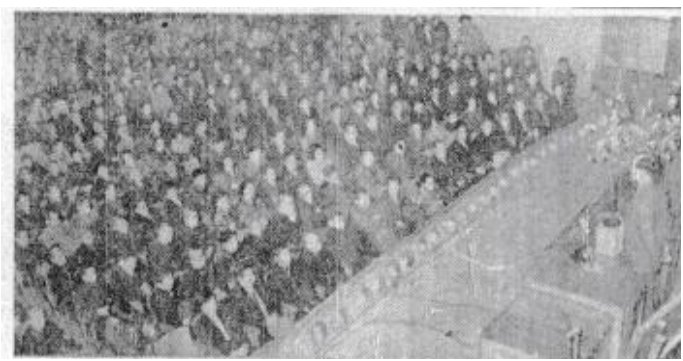

図17　1954 年 3 月 22 日高知市中央公民館で「科学者としての体験について」講演の湯川秀樹博士[47].

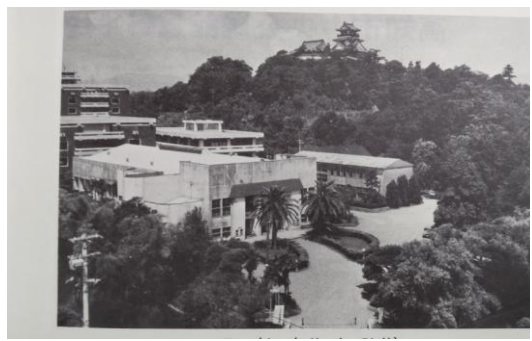

図18　湯川秀樹が講演した高知城真下の高知市中央公民館(左側の本館).　右側は別館. 聴衆は館外にまであふれた[49](今の丸の内緑地).

していた野市町に近い日本3大鍾乳洞のひとつである龍河洞の見物を行っている.

　3月22日午後6時からの一般向け講演（図16）[48]で湯川がどのような講演をするか, 原子力・水爆問題に言及するかが注目された. 図16の新聞社の講演案内には演題がない. 講演は高知市中心部で高知城の真下にある高知市中央公民館（図17, 図18）で行われた. 講演翌日の新聞によると[47]講演の演題は「科学者としての体験について」とある. 一方, 中央公民館の記録には演題は「科学者の立場」[50]とある. 演題がなくても聴衆は湯川の講演につめかけた. 日本人として初めてのノーベル賞受賞者というだけでなく, 高知県ではビキニ海域での遠洋漁業従事者が多くいて, アメリカの水爆実験の被害・影響について原子力の権威がどのような説明をしてくれるのか全国的・全県的にも関心が高かったと思われる. 翌日の新聞[47]は講演について次のように伝えている.「早くからつめかけた聴衆は会場前に列をつくり, 約二千の人々が参集, ＜中略＞開会あいさつにひきつづき澄子夫人同伴で来場した博士の『科学者としての体験について』と題する講演が始められた. 聴講券を手に入れることの出来なかった人たち約三百名も会場北側に設けられた場外聴取席に集まって夜気にもめげずマイクから流れる博士の声に熱心に耳を傾けるという姿もみられ, 人類を進歩に導くか破滅に落すかの岐路に立たされた科学者の宿命, 苦悩あるいは研究の喜びを語る人間湯川博士の講演会は多くの意義を残して午後八時幕を閉じた.」これを書いた新聞記者は「人類を進歩に導くか破滅に落すかの岐路に立たされた」科学者の苦悩についてふれている. 湯川は原子力・水爆問題についてなんら直接的にふれていない. 高知新聞3月25日[51]の講演内容の要約をみると, 講演は原子力のはなしでなく極めて一般的なはなしとして,「科学者の知識欲によって発見されることが人類にとって望ましいことだけでなく, 望ましくないこともある・・・そういう面で科学者は利己主義だとお叱りをうけるかも知れないが, そういうお叱りは甘受する.」[47]と述べている. それでも「原子力」という言葉が6カ所はでて来る. 湯川が水爆問題を意識せざるをえず, きわめて間接ながらもさけることができているのを示している.「原子力」という言葉を含む発言には次のような発言がある.「私が過去五年間くらいアメリカにいて日本へ帰った時, 私から何か原子力の話をきこうとし, 私がそれに触れると権威があるように思われているが, そうではないようである. 私以上に原子力にくわしい人達は沢山いるのである. アインシュタイン博士は偉大な物理学者であるから原子力にくわしいか？そうでもない. 興味を持っている方向も全然違うのである.」. さらに次のような強いメッセージを含む発言もある.「こんな状態であるからやめようにもやめられないのが学問である. 原子力というものは原理的には現在わかっている. 利己的で無責任のようであるが原子力の問題は外の人にまかせておき私は本来的な傾向をおし通し自分のやりたいことに向かって進みたい.」[47]（下線は筆者）

　3月23日の高知新聞のコラム「小社会」[52]は湯川博士の講演について「第五福竜丸の"死の灰"事件以来, 原子力問題は国民の身近な問題となってあらわれている折, 博士にたいする国民の関心は深いが『私はみながいうほど偉い人でもなければ立派な人間でもない』と語る謙虚な博士の科学者としての体験談こそ, 県民の心耳には大きな響きを与えるものであろう.」とし, 湯川を原子力の権威としてとらえている.

　湯川は翌日23日午後, 高知市中央公民館で学術関係者むけの専門的講演を行った. 新聞記者は3月24日にこれについて次のように書いている[53].「学術講演に臨んだ博士は『物質構成は従来永久不変のものといわれていたが, 素粒子の発見によりこれまでのような研究の範囲では解決出来ず, 非局所場ともいうべき無限の場を設けねばならぬ』と述べ, さらに原水爆の将来について『素粒子の性質如何によってこれまでと異なったものへ成長することになろう. 私は原子を平和の子としてあくまで育てる決心だ』と語った.」新聞記者や社会は湯川を「原子力の権威」としてみ, 水爆問題にたいする「直接的な言」を求めているが, 湯川はあくまでそれにふれていない.

　湯川は弟子で中間子論建設の協力者でもある小林稔（京都大学名誉教授）(1908(明治41)-2001(平成13)), 先輩の原子核実験物理学者, 荒勝文策(1890(明治23)-1973(昭和48))らと戦時中に海軍の要請で日本の原爆開発に消極的ながらもかかわらざるを得ない状況におかれたことがあり[54], 原爆や核兵器についてはむろん素人というわけではない. 沈黙をまもった背景には戦前の反省の気持ちも綴った敗戦後間もない1945年11月『週刊朝日』に載った敗戦後はじめての文書「静かに思う」[55]に一端をうかがい知れるかもしれない. この「静かに思う」は翌年, 本『自然と理性』[56]に採録されるが, 採録された「静かに思う」には最後に「付記」が記され次のようにある.「付記　終戦後二カ月ほどの間、色々な新聞や雑誌からの原稿の依頼を固くお辞わ

りして沈思と反省の日々を送って来た．その間に少し気分が落着いて来たので初めて筆を執ったのがこの一篇である．一年後の今日から見るとまだまだ反省が足りないが，その時の気持がある程度まで現われていると思われるので採録することにした． 1946 年 11 月」[57]．

## 8 水爆・原子力ときびしい世論：苦悩と人生の転機

湯川は初めての高知訪問を澄子夫人とともに楽しんだ．高知訪問について湯川は出発 1 カ月前の 1954 年 2 月 28 日京都で「恩師故森総之助元三高校長に物理の手ほどきを受けたので高知は特になつかしい，妻も高知行きを希望しているので出来れば同伴，桂浜などの景勝地を見物したい．」[58]と語っている．そして実際に 3 月 22 日午後に龍河洞を訪ねている．その日の晩一般市民むけ講演終了後，3 月 22 日夜，には湯川夫妻は高知三高同窓会に出席し旧交を温めている．この親睦会は公式に報道されているが，それへの出席については，湯川夫人が翌 3 月 23 日午前訪ねた観光名所桂浜で取材の新聞記者にもらしている．「ほんとに高知はいいところですね　人情も景色も！　―それに酒もイイレ，みなさんお強いのにはおどろいた，酒も少しなら頭脳の刺激になっていいもんですヨ」[53]．（若いころお酌をしたことがあり，小説のモデルにもなっているとされる祇園の女将に聞き「湯川さんは強かったですよ」とかつて聞いたことを思い出した）．エンドレスな酒盃の交換が続く土佐流の三高同窓の酒宴を楽しんだようである．この桂浜の清遊は中学時代の親友大岡高知赤十字病院長夫妻が案内している．院長の案内で湯川は夜須小学校 PTA の自主的運動で建立される「湯川胸像」のさきがけとして県民の自主的運動で建立された太平洋を望む志士「坂本竜馬銅像」の勇士[59]を見上げたことであろう．高知での酒について湯川夫人は同日午後一時半からホテル三翠園で開かれた「湯川夫人を囲む座談会」に出席し，さらに語っている．「『ご主人への心づくしの秘ケツは？』との質問に『頭を休ませるのが第一だと思ってこのごろはお酒を飲ませております』と答え，三たび満場を爆笑させた」[53]．湯川は念願の土佐闘犬を見たいと思っていた．湯川のために，3 月 23 日最後の講演であるは高知市中央公民館での午後 1 時半からの学術講演会で講演を終えた後 3 時に会場前広場で高知市・升形の斎藤秀吉所有の土佐闘犬・美濃号が披露され，湯川はおおいによろこんだ．高知新聞の伝えるところによると『『土佐に来たからにはぜひ闘犬とやらを一目！』という博士のたっての希望によるものといわれ初めて見る犬の化粧まわし姿に『ずいぶん大きいネ』ビックリしていた」とある[53]．午後 3 時 58 分には国鉄高知駅から鉄路京都への帰路についた．

湯川は高知滞在中に水爆・原子力問題に直接ふれることを専門外のことだと断固拒否し通した．湯川が高知駅を離れた 23 日の新聞夕刊に『湯川さん』と題する署名入り[60]の次の記事が載った[61]．「湯川博士が来高した．ノーベル賞の湯川といえばこどもでも知っているせいか，高知駅は大した歓迎ぶりだった．<u>ちょうどビキニの第五福竜丸事件で原子力への関心が一段と高まっている折だけに</u>，原子物理学の権威，湯川博士の顔を一目でもみたいという・・・もかなり多かったようである．<u>湯川さんは肝心の原爆問題については完全にノーコメントでおし通した</u>．記者団がしつこく質問すると<u>『私は原子力については何も知らない．私のやっている学問はそんなものではない』</u>と，いささか迷惑そうだった．」この記事はさらに次のように書いている．「<u>湯川さんの原爆ノーコメントは何もいまにはじまったことではない</u>．昭和二十四年に「中間子理論」でノーベル賞をもらったとき，訪れた UP の記者にも<u>『私は原爆については話したくない．私たちの話題はメソン（中間子）だ』</u>と語っている．しかし一般の素人考えでは米国のラビー教授[1]（一九四四年度ノーベル物理学賞）が『湯川博士の新学説は第二次大戦に先立つ十年間の基本原子学説にもっとも重要な寄与をなすものだ』といっているとおり，<u>ひろく原子力問題の最高の学者とみている</u>．湯川さんの沈黙は事をいやしくもせぬ学者的良心と解すべきであろうか．こんどの来高を機会に香美郡夜須小学校では博士の胸像除幕式を行うそうである．夜須小のこどもたちは湯川さんをどう考えているのだろう．恐ろしい原爆の恐怖を除いて原子力を平和へみちびいてくれる人，そんな漠然とした期待が学園の胸像となったのではないか．<u>湯川さんがいくらノーコメントをつづけても国民は決して「湯川さんと原子力」を切り離して考えないだろう．</u>」（下線は筆者）

---

[1] I. I. Rabi （1898-1988）[62]

　原子力問題，水爆問題に答えない湯川にたいするかなりてきびしい記事，世論の声である．「恐ろしい原爆の恐怖を除いて原子力を平和へみちびいてくれる人，そんな漠然とした期待が学園の胸像となったのではないか」の言説は「湯川胸像」の設立趣旨とまったく離れている．湯川胸像建立は前年 1953 年秋に企画され，胸像は水爆実験以前に完成されており，建立の設立趣意書には「先生の偉勲を讃へると共に児童をして先生の風貌を景仰し其の偉業を感謝し深く科学研究の重要性を感銘せしめんがため胸像建設を計画する・・・感激に堪へず仍て健像の由来を刻して之を不朽に傳へ深甚の感謝の意を表す」[7]とある．ただ，記事は県民・市民が水爆による被爆に恐れおののき湯川さんにすがり助けを求めている気持ちを代弁したのであろう．湯川はこの記事を読むことなく夜須小学校の除幕式にのぞみ，4 つの講演をおこなった．湯川胸像は突如沸騰してきた時世の水爆・原子力で世論や新聞からはかすんでしまう．湯川胸像はその後新聞記事とされることも話題になることもなかった，湯川も以後高知訪問のことは語らず記録に残さなかった．湯川胸像は経年とともにわすれさられ，歴史の激流に埋もれた．

　帰洛後，湯川はこれまでの学究生活から「平和を探求する」「核兵器廃絶運動の実践者」としての物理学者に変わる．1954 年 3 月高知を訪問した湯川はこれまでつらぬいてきた研究一筋の「自己本位」の物理学者としての学究生活・学者人生と世論の求める「原子力問題の最高の学者」とのはざまにおかれ揺れ動き人生の転換点におかれた湯川であった．

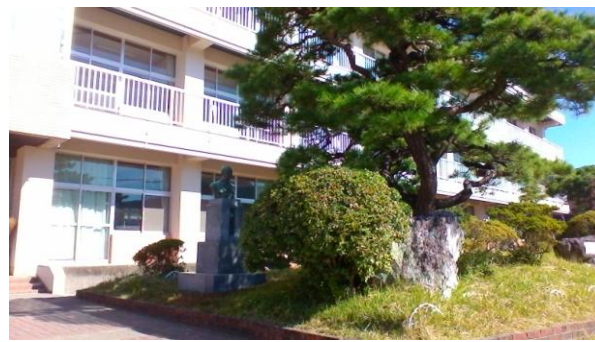

図 19　大きく成長した松のもとにたたずむ夜須小学校改築記念園の湯川秀樹胸像．2018 年 11 月 4 日筆者撮影．

　湯川は中間子理論をつくった戦前の大阪帝大時代も，1939 年京都帝大にうつり中間子論を発展させる研究においても，1948 年から 5 年間のアメリカ滞在中も，そして 1953 年夏の帰国，京都大学教授復帰から高知訪問直前まで，一貫して「自己本位」的，「利己的」研究者・物理学者[51]として真理の探究，素粒子の本質・時空の構造の探求にひたすら没入していた．湯川は 3 月 22 日夕方の一般市民むけ講演でこう断言している．（わたしは）「何かやると徹底的にしたい性分で骨を折って自分の力でしていきたいという本来的な傾向が幼い頃から今日まで続いている．私は大した才能はもっていないが自分でいいと思ったこと，自分でやりたいことだけしかやらない．自分のやっていることをとやかくいわれるのも嫌だし人のことを批判するのも嫌だ．要するに大へん利己主義でケシカラン人間だ．」[51]（下線は筆者）．湯川の講演内容を報じた新聞は見出しに「研究は知識欲のため　　利己主義のそしりを甘受」と 4 段抜きの大きな見出しを付けている[51]．高知訪問を契機に，このふつうの研究生活・学者人生から「人生観がかわる」転機に遭遇するとは思いもしないことであったろう．寺田寅彦も湯川秀樹も若いころ読んだであろう『科学と方法』で著者フランスの理論物理学者ポアンカレー（H. Poincaré 1854（嘉永 7）−1912（明治 45））はこう言っている[64]．「科学者は実益がある故に自然を研究するのではない．自然に愉悦を感ずればこそこれを研究し，また自然が美しければこそこれに愉悦を感ずるのである．自然が美しくなかったならば，自然は労して知るだけの価値がないであろう．（略）科学者はこの美のために，おそらくは人類の将来の幸福のためよりもむしろこの美のためにこそ，長い苦しい研究に身をささげるのである．」寺田寅彦は自らこれを翻訳するとともに[65]，その著『科学者と芸術家』[66]で同じことを言っている．学問の面白さ・物理学研究の尽きない奥深さに魅せられ大学退官前も後も「原子核の $\alpha$ 粒子構造と原子核の虹」を追い求める[67]筆者も「大へん利己主義でケシカラン人間」となるだろうか．自然と対峙の研究は魂の籠る執念である．

## 9　決断：湯川は清水の舞台から飛び降りた

　湯川は 3 月 23 日最後の学術講演を終え午後 4 時前に国鉄高知駅をたち帰路についた．当時の列車は蒸気機関車であまたのトンネルをくぐりぬけ四国山脈を越えて高松にいたるだけでも 5 時間かかる京都までの長い道中であった．湯川は車中，高知駅での離別のさいに地元の小学生から贈られた詩集を読んだ．車窓の四国の春をながめつつ，滞在中のいろいろことに思いをめぐらしたことであろう．ビキニ水爆で日本漁船員が放射能被

爆し火傷をおい，マグロが放射能汚染で食べられなくなり，日本中が大騒ぎになっていた．湯川は高知では皇室が定宿とし清流・鏡川がながれ坂本竜馬も泳ぎ筆者も少年時代もそうした上町の竜馬生誕の地のとなりにある老舗の城西館に宿泊していた[20,32]．先に引用したように，高知へ出発する直前，湯川は日記に「3 月 16 日 火 曇 ＜略＞三月一日ビキニ環礁北東約百マイルの地点で水爆実験による真っ白な灰を被ったマグロ漁船第五福竜丸帰港、火傷の傷害を受けた乗組員を診断 水爆症と推定」と書き，ビキニ水爆事件に強い関心をもち，高知滞在中もそのなりゆきを気にかけ追っていたはずである．老舗ホテルでもいつものように新聞をよんだことだろう．3 月 22 日の湯川講演を報じた翌 23 日の高知新聞朝刊，図 17 の真下には，ビキニ水爆の記事がのっている．

高知滞在中の湯川には新聞記者が同行し，「土佐犬を見た」など湯川の一言一句を注視している．ビキニ水爆について特ダネを湯川から引き出そうとする記者からも，湯川はビキニ水爆の世論の風圧を感じていたことであろう．「科学者としての体験について」という題の一般向け講演で，いかに自分が「原子力」と無関係であるかを論証しようとして，「原子力」という言葉を何度も自らの口から発した．湯川はすでに「原子力」からのがれることができなくなっていた．一般民衆にとってその時期，「原子力」という言葉は「ビキニ水爆」と不可分的にむすびついている．湯川は高知滞在中でも帰洛途上でも思索をめぐらしたことであろう．

湯川は京都に帰り生涯の転機となる決断を下す．すくなくとも 3 月 28 日までに決断している．そして湯川は 3 月 28 日清水の舞台から飛び降りる決心をした．その苦悩の決断の序章は高知滞在中から始まっていたとみることができよう．高知を去り 1 週間後，3 月 28 日「原子力と人類の転換」を執筆する．「原子力」にたいするはじめての「社会的発言」である．湯川の声明文は 3 月 30 日，毎日新聞[68](図 20)に特別寄稿「原子力と人類の転換」として掲載された．世紀の「声明」であり，これは，原子力の権威としてみずから社会に登場する学者・湯川にとって「人生の転機」であった．湯川は自分の原稿が新聞紙面に載ったことを確かめている．湯川は日記に残している[45]．「3 月 28 日 日 雨 家に居て 毎日新聞原稿「原子力と人類の転機」＜略＞ 3 月 30 日 火 晴 毎日」朝刊に「原子力と人類の転換」第一面に出ている＜略＞ 4 月 2 日 金 晴 ＜略＞午後 国会 自由党総務会で 原子力について話す」．

日記[45]には高知訪問についての記述・言及はまったくない．水爆と人類の危機に思索を集中する湯川には胸

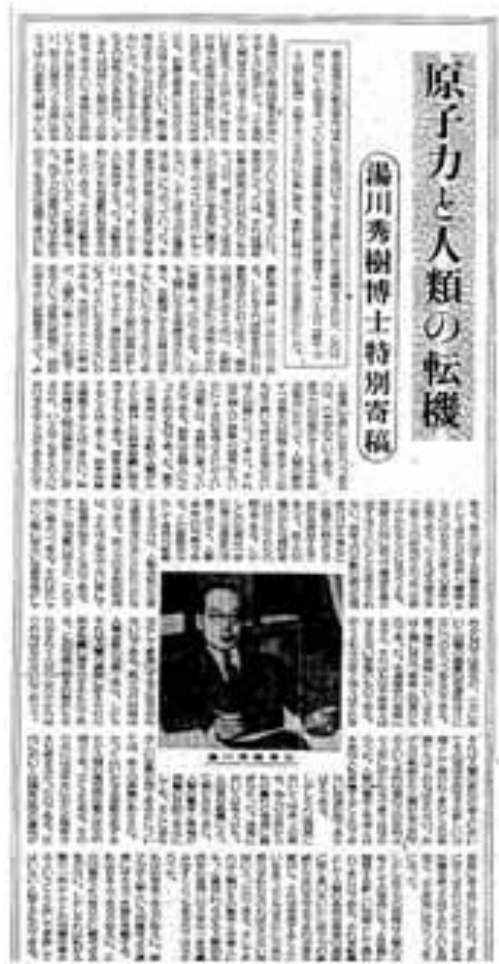

図 20 1954 年 3 月 30 日毎日新聞朝刊には発表された湯川秀樹博士の原子力にたいするはじめての社会的発言である「原子力と人類の転換」の記事[68]．

像除幕式へのはじめての高知訪問など個人的なことを日記にしるす心境でなかったのであろうか．日記の記述や3月21日夕刻高知駅到着時での記者会見での発言，（原子力については）「ご承知のように一般に言明したことがなく全く関知しないところで私の研究外だ。この問題は答えられない。」の発言からも，この湯川声明は毎日新聞社に依頼されて書いたというより，湯川がみずからの意思で発表したのではないか．日記には毎日新聞社に依頼されたとの記述はない．日記にみずから「『毎日』朝刊に『原子力と人類の転機』第一面に出ている」と記し，新聞掲載を確かめたのも，うなずける．しかし自発

的か依頼か，いずれであるかはまったく重要ではない．

　3月21日高知での記者会見から3月28日毎日新聞原稿の執筆の7日間に湯川の人生に「生涯の転換」が起こった．水など物質が固体から液体へとかわる形相の劇的な質的変化である「相転移」，おそらく高知滞在中からその「相転移」は始まっていたであろう．3月22日の一般市民向け講演[51]で「科学者の体験について」という漠然とした演題を設定し，アメリカでの研究生活やアインシュタイン(A. Einstein 1879–1955)[69]の話にふれ，研究者がいかに自己本位，興味本位な人間であるか語り，社会や世の問題といかに隔絶していることを強調し，水爆・原子力には直接ふれず，原子力とは無関係だと話す論理展開には湯川の「苦悩」が如実に映されている．3月28日に「人類の転機」を執筆するまで，その苦悩は続いたであろう．「人類の転機」はその一員たる「湯川の転機」でもあった．

　湯川のはじめての社会的発言がいかに大きかったか．湯川は日記に「4 月 2 日 金 晴 ＜略＞午後 国会自由党総務会で 原子力について話す」と記すように，社会・政治においても重要な存在となっていく．

　湯川は約 2 週間後さらにふみこんだ「談話」を発表する．朝日新聞は 4 月 16 日夕刊で「『原爆問題』に私は訴える ＝もう黙ってはいられない＝」との見出しで湯川の談話を報じた[70]（下線は筆者）．ビキニ水爆実験で「日本人漁夫が放射能による火傷をうけマグロが食べられなくなって以来一カ月」ということで，AP 通信社のアンソン東京支局長が「4 月 15 日京都大学湯川記念館に湯川秀樹博士を訪問し，＜略＞博士の見解をたたいてみた．以下その談話の要旨である」と書く記事である．その湯川談話を引くと，

　　「日本人は，最近のビキニ近くで行われた水爆実験によって大きな衝撃をうけた．＜略＞私は日本ばかりでなく，他のどこにおいても，政治運動からは遠ざかるよう，あらゆる努力を重ねて来た．しかし，こんどの問題は原子力対人類の問題である．＜略＞ 私は，こうした複雑な問題からは遠退いていたいと思っていた．しかし，私は，一個の人間として責任を感じる．日本人が，もし実験の時期と危険の広さとについてもっとよく知らされていたならば，大いに助かっていたことであろう．＜略＞われわれは，米国ばかりでなく，ソ連，英国その他の強国に対してもこのことを訴える．水爆実験で火傷をうけた漁夫諸君の容体が快方に向かっていると聞いて喜んでいる．＜略＞」[70]（下線は筆者）

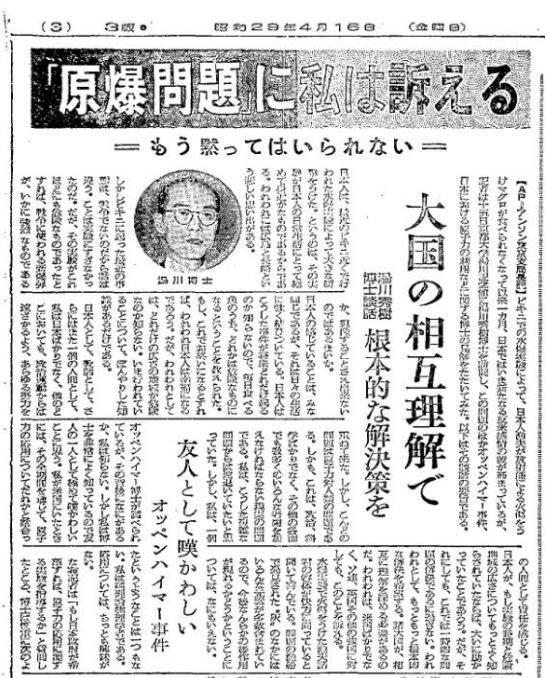

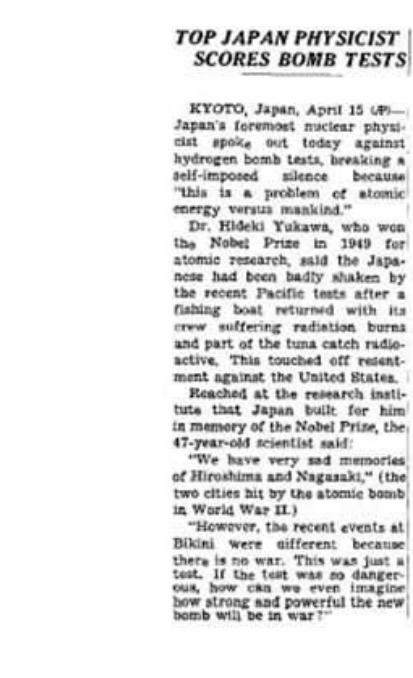

図21　朝日新聞(1954 年 4 月 16 日)掲載の湯川秀樹の談話：「原爆問題」に私は訴える ＝もう黙ってはいられない ＝[70].

図22　新聞ニューヨーク・タイムズ(1954 年 4 月 16 日)が報じる湯川のビキニ水爆実験にかんする談話[71].

　アメリカの新聞ニューヨーク・タイムズは 4 月 16 日「日本のトップ科学者水爆実験を非難」（「TOP JAPAN PHYSICIST SCORES BOMB TESTS」）と即座に湯川発言を報じた[71]．（英語版ニュースについては 3 月 30 日の毎日新聞に出した湯川声明も翌 3 月 31 日に英字新聞「The Mainichi」に「Atomic　Energy and Turning Point for Mankind」として報道されている）．湯川の発言は国内だけでなく世界からも注目され，その発言の意味・影響力は非常に大きかった．湯川はその大きさを知っているからこそ，断固かかわることを拒否して続けてきた．ニューヨーク・タイムズは次のように報じている[71]．

　4月15日日本の京都．<u>日本の最高の核物理学者がきょう，水爆実験に反対を表明した．自身に課してきた沈黙を破ったのは「水爆実験は原子力対人類の問題だ」という理由である．原子研究(atomic research)</u>で1949年ノーベル賞受賞の湯川秀樹博士は最近の太平洋での水爆実験で，操業していた漁船の船員が，放射能による火傷を負いまた漁獲したマグロが放射能汚染され帰ってきたことに日本人は大きな衝撃をうけていると語った．これは米国にたいする憤りを引き起こした．ノーベル賞受賞を記念して日本政府が建てた研究所にもどった47歳の科学者は「われわれにはヒロシマとナガサキという非常に悲しい思い出がある（この二つの市には第2次世界大戦で原子爆弾が落とされた）．しかし，ビキニで起こった最近の事態は，戦争ではないのだから話は違う．だが，その実験がこれほどにも危険であったとすれば，戦争で使われる新爆弾がいかに強力なものであるか，想像することすら出来ないのではあるまいか」と語った．（下線は筆者）

　この間の湯川の原子力に対する態度の急激な変化は次のようにまとめられる．「超然」（3月21日高知記者会見：湯川胸像除幕前）「苦悩」（3月22日高知講演：湯川胸像除幕後）→「決断」（3月28日毎日新聞声明「原子力と人類の危機」執筆）(4月16日湯川談話「『原爆問題』に<u>私は訴える　＝もう黙ってはいられない</u>」）．

# 10　『しばしの幸』出版と「原子力と人類の転機」

　1938年四国を初めて訪問した湯川は随筆「四国の秋」を残した．1954年の旅で湯川は高知訪問から帰ると読売新聞社の依頼で『しばしの幸』[73]を本として出版する．その序文は4月6日に書かれている．『しばしの幸』には第I部に「京都の秋，イタリアの夏，スエーデンの冬，アメリカの春，そこ冷え，サービス，ダンスのできない人間，想像力，記憶，北海道の夏，四国の秋，歳末の感」，第II部には，「旅のノートから，ロボット，ハドソン河畔の秋，贅沢にならない快適な生活，日本のお祭り，祇園祭の印象，日本のお正月，変るもの変らぬもの，ノーベル賞を受けて」，第III部には，「誤解と弁解，甘さと辛さ，科学が生かされるということ，徹底ということ，歳月，一科学者の人生観，東洋と西洋，東洋的思考，模倣と独創」，第IV部には，「少年時代の読書，読書漫録，プリンストンだより，仁科芳雄先生の思い出，長岡半太郎先生のことなど」，第V部には，「物理学の二十年，暗中模索，理論物理学の現状，物理学の前途，原子力と人類の転換」．学究生活における折々の思い，「しばしの幸」が記されている．「四国の秋」には1938年湯川が家族総出ではじめて見知らぬ四国を訪ねたときの思い出が記されている．

　高知訪問は湯川にとって京都一中・三高の級友にもあい，夫人と初めての景勝の桂浜・龍河洞をたずね[74]，また夫人は「湯川夫人を囲む座談会」で坂東流の舞を披露し[75]，たのしい旅であった．夫人は帰洛後，高知駅での見送りで詩集をくれた小学6年生（小坂坂小学校，偶然か筆者はこの母の母校に後年4年生の時に（1957年秋）うつる）あてに礼状を書いている[76]．過去の湯川日記を見ると湯川は日記にしばしば家族のこと，個

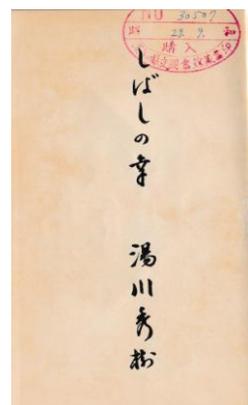

図 23　湯川秀樹著『しばしの幸』(初版)
[73]．高知県立図書館蔵．書名は湯川の直筆．序文は 1954 年 4 月 6 日，湯川胸像除幕式の高知訪問後間もなくに書かれ，最終篇に「原子力と人類の転換」が収められた．

人的なことも記している[77, 78]．また，各地に旅行すると和歌を詠み，四国の旅では伊予の道後温泉や徳島での和歌が知られている[79]．だが高知訪問のあと湯川による随筆「四国の春」は書かれなかったし和歌も残されなかった[80]．ビキニ水爆の第五福竜丸事件に接し，湯川は高知で人生観の転機となる原爆・水爆，「原子力」について社会的発言をするべく強い社会的圧力うけている．その決意をした可能性さえありうる．本『しばしの幸』の最後の一篇に随筆「四国の春」や和歌があれば，素晴らしい『幸』の完結である．だが湯川は最終の一篇に「幸」についての話題でも随筆でもなく，「しばしの幸」でもない，核兵器に関する所見「原子力と人類の転換」を収めた．最後の一篇が形式においてまた内容的においても「随筆」でなく異質な存在であることは明らかである．のちに編集される湯川著作集においてに，「原子力と人類の転換」は「静かに思う」とともに第5巻『平和への探求』[81]に，後に書かれた「世界連邦への道」「核時代の平和思想」などとともに収められている．湯川の並みならぬ，「転換」の決意をうかがわせるものである．

湯川は，高知滞在中あるいは帰ってから，3月28日に書く「声明」の構想を練ることにかかわり，夜桜町除幕式の思い出，龍河洞の思い出，同窓会の思い出，桂浜の思い出，闘犬見物の思い出，初めての高知の思い出，などを記す余裕がない，その後の生き方を変えることになる大きな「人生の転換点」にいた．4月16日の朝日新聞にでる談話「『原爆問題』に私は訴える ＝もう黙ってはいられない」で湯川の転機は決定的になった．湯川の転機は歴史と時代における避けがたい宿命でもあった．湯川は核廃絶の運動にむけて歩みだす．湯川の弟子の武谷三男（1911（明治44）-2000（平成12））[82]など素粒子論・原子核理論などの理論物理学者，素粒子論グループの科学者もそうであった．原子核の世界を知りその研究にかかわった理論物理学者の歴史的な宿命であった．しかし湯川は単なる核兵器廃絶の運動の実践者にとどまらなかった．湯川が好んだ荘子の『荘子』内篇逍遥游[83]をおもわせるように，スケールの大きい理想主義的な包括的な平和の創造のため世界連邦運動」[84]で世界連邦世界協会会長にもおさまり，夫人[11]とともに核兵器のない平和の実現をめざす．

## 11 1954年高知訪問のころの湯川の学問と研究

湯川が「自己本位」的に没入していた研究とはどんなものか．湯川の「人生の転換点」となる1954年はじめ前後のころの湯川の学問・研究を見ておく必要がある．湯川は1953年7月アメリカから帰国後，7月1日付でコロンビア大学から京都大学教授に戻り，8月1日には湯川のノーベル賞受賞を記念して京都大学に設立された日本で初めての全国共同利用研究所，湯川記念館・基礎物理学研究所の所長に就任，9月1日には基礎

| |
|---|
| 1. 山本隆男(弘前大)高スピンの解析的表示について |
| 2. 町田茂，井上昭太郎(広島大理) 素粒子の角運動量について |
| 3. 田中正(京大理)エネルギーのずれを伴う定常状態について |
| 4. 喜多秀次 (京大理) Heisenberg 運動方程式の解について |
| 5. 林忠四郎(浪速大工)非局所的相互作用の場の Hamiltonian 形式 |
| 6. 勝守寛(京大工)非局所的な直接相互作用による二粒子の束縛状態 |
| 7. 濱口實 (京大理)運動学的に広げられた素粒子の運動論的解釈 |
| 8. 江夏弘(京大理) 相対論的固有値問題としての素粒子の質量スペクトル |
| 9. **湯川秀樹(京大理)非局所場について** |
| 1. 福田信之 （東京教育大） π中間子 |
| 2. 佐々木宗男(都立大)高木修二，谷純男(京大)核力 |
| 3. 吉田思郎 （基研） 原子核の構造 |
| 4. 長谷川博一 （大阪市大） 原子核の飽和性 |
| 5. 小川修三 名大)Comment; Tamm-Dancoff の方法と核力の飽和性 |
| 6. 早川幸男 基研)木下，南部の新らしい論文の紹介 |
| 7. 澤田克郎 （東大） 核子のまわりの中間子雲 |
| 8. **内山龍雄 （阪大)Non-local Interaction について** |
| 9. 林忠四郎 （浪大） Comment |
| 10. **湯川秀樹(基研) Non-local field について** |
| 11. 江夏弘 （京大） Comment（Mass Spectrum） |
| 12. 早川幸男 （基研） 新粒子の実験データ |

図24 湯川秀樹が口頭発表した1953年10月の日本物理学会第8回年会(東大教養学部)(10月15- 20日)，基礎理論セッション(10月17日午前)のプログラム[85]．

図25 湯川秀樹が口頭発表した1954年の1月京都大学基礎物理学研究所「素粒子論シンポジウム」(1月15- 16日)のプログラム[86]．

物理学研究所教授と配置換えされ，理学部物理教室には併任教授となる．1953年9月11日－24日まで湯川は日本で敗戦後初めて開かれる国際会議である理論物理学国際会議（会場は京都と東京）の会長として，京都と東京で会議を成功裏に導く大役を果たす．海外から55名参加し，ノーベル賞受賞者はその後の受賞も含めると17名で，湯川の世界的の評価の高さを物語っている．10月には日本物理学会第8回年会（1953年10月15日－20日，東大教養学部）の基礎理論セッション（10月17日午前）[85]で9番目の最後の登壇者として「非局所場について」と題して講演している．図24は1953年秋ごろの学問研究状況が理解できる講演発表プログラムである．

　湯川は高知へ出かけるすこし前の1954年1月におこなわれた京都大学基礎物理学研究所での「素粒子論シンポジウム」[86]において，10番目の講演者として「non-local field について」と題して前年秋と同じ題目で口頭発表している．図25はプログラムである．この研究会「非局所場理論」は昭和28年度に始まった基礎物理学研究所の最初の長期研究計画である[87]．（注目されるのは湯川が出席しているなか，8番目に内山龍雄による発表が行われていることである[88]）．湯川は非局所場理論の研究に集中し，研究人生の後半ひろがりのある素粒子と時空の問題へと研究を進める．

　湯川の研究スタイルは中間子論発見のときとおなじくスケールの大きい「哲学的な背景の上での物理」[89]として広がりのある素粒子像を求め非局所場の理論に没入し，壮大な理論の構築をめざしていた．それは1968年ごろまでに「素領域理論」として数学的にも美しい理論として定式化される[93-96]．素領域理論を構築した湯川は学問的にも研究人生においても高い境地に達していた．渡辺慧[98]は「天才の仕事」といっている．

## 12　核廃絶・平和主義の実践者として湯川の高知再訪

　湯川には，1954年高知訪問時の高知駅到着での記者会見で水爆実験にたいする記者の冒頭質問に答えず「ノーコメント」をつらぬき，世論・県民・国民が渇望している水爆・原子力問題について講演でも一切ふれなかったことに忸怩たる思いがあったのか．生涯誠実な生きかたを貫いた湯川は京都大学を定年退職する前の年である1969年6月21日，夫人とともに再度高知を訪問した．理論物理学者としてだけでなく世界連邦による核兵器のない包括的な平和の創造をめざす世界連邦の実践者としての再訪である．湯川は1954年の高知訪問から帰洛後間もなく毎日新聞にだした声明発表後は，以前にもまして理論物理研究の権威者としてだけでなく，原子力の権威として，その言動がマスメディアの注目の人となっていた[99]．

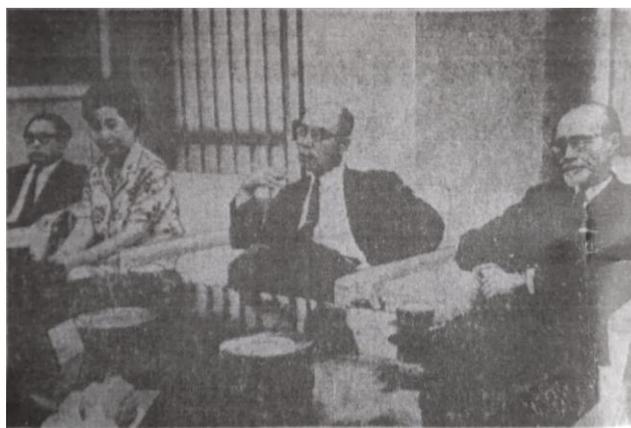

**図26　1969年『核時代と平和』の講演のため高知県南国市と高知市を訪れた湯川秀樹(中央)と夫人湯川スミ(左)．右は谷川徹三　[102].**

　高知市では核兵器廃絶と平和の創造の問題で講演し[100]，「宿題」に真正面から答える．科学者湯川の誠実な真摯な姿勢がつらぬかれている．6月22日午後，高知市での世界連邦平和講演会で湯川秀樹は「平和の創造」と題し，元法政大学総長・哲学者の谷川徹三（演題は「世界政府の理念」）とともに講演する．[101]．湯川澄子から湯川スミと改名した夫人も「みんなが幸福になるために」と題し冒頭講演した．

　同じ日の午前中に，湯川は東に隣接する南国市を訪れていた．湯川の今回の高知訪問は市制10周年を迎えた南国市の招きによるものである[102]．湯川胸像が立つ香南市と高知市にはさまれた南国市は筆者が4年生前頃までの幼少期（十市小学校，父の母校でもある）をすごしたところでもあり感慨深い．湯川は和歌を好んだが，この地はかつて日本文学史に新しい境地を開く『土佐日記』[103]を残した歌人・紀貫之（872(貞観14)-945(天慶8)）が国司として滞在し，また京への帰路荒天で避難し正月を過ごした大湊のあるところでもある．湯川先生は中国の思想家・荘子の哲学[83,104]とともに真言密教の祖である天才思想家・空海[105]を好いたが，住職が

紀貫之に餞別を贈ったと『土佐日記』(12月24日)にしるされる四国88カ所28番札所・国分寺や太平洋をのぞむ31番札所・禅師峰寺がある歴史の由緒のあるところである．幕末の志士・武市半平太の生家(高知市仁井田)も近い．荘子や空海を好いた湯川のスケールの大きい思想は学生時代にきいた講義や核廃絶の平和主義の運動にも滲むように出ていたように思う．その南国市では湯川と湯川先生にあこがれ高知県立東工業高校で，湯川は「核時代と平和」，湯川スミ夫人は「みんなが幸福になるために」と題し講演を行っている[106]．湯川は『「人類のつくった絶対悪は核兵器だけだ．国家相互のにくしみや不信を取り払い，友愛と信頼の関係を築くことによって，地球上から核兵器をなくすことが平和をつくりだすことである』と高知での講演を締めくくった」[107]．1954年3月の高知駅冒頭記者会見での「ノーコメント」から15年，湯川は夫人とともに高知の地に堂々と立ち，県民にたいし原子力，核兵器問題についてみずからの核廃絶運動の実践を踏まえ講演．いかにも湯川らしい堂々たる誠実な物理学者の風格である．1954年記者会見で質問した新聞記者はこの講演をどう聞いただろうか．この湯川講演を聞いた聴衆のなかには後にあきらかになる室戸市や土佐清水市・宿毛市など，ビキニ水爆で被爆した高知県の遠洋漁業者の関係者がいたかもしれない．

このころの湯川先生を筆者はよく覚えている．哲人を思わせるような風格をそなえた理論物理学者であった．筆者は湯川先生が1954年，筆者の生まれた地・高知に来られ，しかも湯川胸像が建立されたことなど，幼小小学1年生で全く知らなかったように思う．だがその後，物理学と湯川先生にあこがれ京都大学物理学科に進み，1969年湯川先生が高知を再訪されるころは京都大学で湯川先生の講義や研究を聞く学部学生そして大学院生・研究者になっていた．1967-1968年には理学部で学部学生として湯川先生の講義「物理学通論」の講義を聞いていた．講義のなかで日本を訪問していたハイゼンベルグの特別講義があったことや，講義の最終で「量子力学の観測理論」[108]を聞いていたことを記憶している．ハイゼンベルグの講義内容は「観測理論」の講義とともに今も手元にあるセピア色になった「湯川教授講義ノート」にみることができる．湯川先生の一人思索するかのよう静かに声を絞り出すように話される講義は耳をすまさないとほとんど聞きがたく，観測理論などは哲学的・深遠で哲人の話を聞くような感があった．北白川の理学部植物園に隣接し閑靜な湯川記念館(基礎物理学研究所)の3階の講義室で聞く湯川先生の講義はいつも静寂でなにか道を究めた最高の哲人の声を聴くような感じであった．このころ湯川先生は学問的にはみずから追いきわめた深遠な素領域理論に到達し，思想的には世界連邦による核のない世界をめざすという平和主義の非常に高い境地に達していたように思う．

核廃絶運動の実践者として高知を訪問した1969年，湯川先生は湯川記念館・基礎物理学研究所の所長であるとともに理学部物理教室の併任教授であった．日常的に基礎物理学研究所の所長室におられたが，理学部物理学科にも部屋はあった．南に面し日当たりがよい部屋で，湯川教授室に隣接しては初期の弟子で中間子論建設にくわわった原子核理論の小林稔教授室があった．筆者は学部では素粒子の卒業研究をおこなったが，そのころ暗黒期にあった素粒子論ではなく原子核理論を専攻し小林先生の研究室に進んだ．（素粒子論を究められた湯川先生も素粒子論でなく生命科学をきりひらくことが重要だと講義やもそれ以外でも学生によく話されていた．）当時物理第2教室はたいへん手狭で通称タコ部屋にいる大学院生の研究場所に困り，素粒子・原子核でも高名な教授室にも大学院生が同室することになっていた．筆者は恐縮しながらも小林教授室に同室させていただき，研究や多方面で小林先生から直接薫陶を受ける幸運に恵まれた．小林先生は湯川先生が1948年招待され滞在しアインシュタインもいたアメリカ・プリンストン大学で研究生活をおくり，ゲージ場理論のヤン・ミルズ理論が発表される歴史的なセミナーにも出ている[91, 92]．小林先生は1954年の湯川高知訪問の10年後，1964年高知市を訪れ，湯川先生が講演したと同じ中央公民館で(8月2日)高知市中央公民館主催の第14回「夏期大学」で「素粒子の世界」[109]と題して講演，このとき高校3年生の筆者はこれを聞いている[110]．（会場にはまだ冷房がなく講師の横に氷柱が立ててあった．）この話を聞き翌年4月京都大学へ進む気持ちがいっそう後押しされた．後年小林研究室に入り，小林先生に話すと大変喜んでくださった．

## 13 おわりに

湯川博士の偉業をたたえるとともに核廃絶・平和への願いをこめられた高知県夜須小学校の湯川胸像(図 1,図 27)は，65年の風雪を経て2019年ふたたび広く世に，世界に知られることとなった[7]．国内，世界に偉大な科学者の銅像や胸像はすくなくないが，没後顕彰して建てられたものが多い．偉業をなした科学者・物理学者

のなかで存命中に建立されご本人が除幕式に出席した胸像はあるだろうか．高知県夜須小学校の湯川胸像は存命中の湯川先生の願いの籠った胸像である．湯川の核廃絶・平和主義の実践者としての生涯の転機はこの湯川胸像除幕の高知訪問から始まった．

中間子論発表から100年を迎える2035年には湯川先生の教えをうけ，じかに接し話を聞いたものはもはや地上にはそれほどいないことだろう．湯川は高知訪問後人生の転機を迎えたと語り，以後核廃絶・平和思想を発展させる[84,111]．「再発見された」湯川胸像[7]が湯川秀樹先生の思いのこもった科学への希望と平和への願いとして記憶されつづけることをのぞみたい．筆者は湯川胸像のことを知っておれば，そのことについて湯川先生にお聞きしたことであろうに，先生が泉下のひととなられてから湯川胸像が故郷にあることを知ったのも何かの因果かもしれない．

### とはにあれの松の茂れる胸像は子どものみらい平和のねがひ

湯川は1981年9月8日京都にて歿する．ビキニの水爆について湯川は高知を訪ねた1954年和歌

### 雨降れば雨に放射能雪積めば雪にもありといふ世をいかに

を残し[79]，広島の平和記念公園の平和の像「若葉」には1966年に詠まれた湯川の和歌が刻まれている．

### まがつびよふたたびここにくるなかれ平和をいのる人のみぞここは

湯川が1954年4月16日の朝日新聞にだした談話「『原爆問題』に私は訴える ＝もう黙ってはいられない」で，「水爆実験で火傷をうけた漁夫諸君の容体が快方に向っていると聞いて喜んでいる」いると案じた「漁夫」は，第五福竜丸以外にもたくさんいたことがその後明らかになる．湯川の人生観をかえ「生涯の転機」となったビキニ水爆実験で被爆した高知県内の漁船員の実相の解明は，高知県宿毛市の高校教諭/山下正寿(1945−)によってはじめられた．宿毛市は足摺岬にも近く古くは武市半平太とも親交があり討幕の戊辰戦争に従軍し，その後は渡米渡英し欧米の学問をまなび，「学問の独立」を唱え官学に対する私学を創立した明治の英傑・小野梓や「日本の独立」を成し遂げた宰相・吉田茂(1878(明治11)−1967(昭和42))など多数の「郷土の偉人」を輩出した．直木賞作家・山本一力は「独立独歩」のまちと記すが，この地にも1954年のビキニ水爆で被爆した漁船員がいた．1985年よりはじめられた地元高校生による調査活動「幡多高校生ゼミナール」によるねばり強い漁船員をさがし，聞き取り調査でしだいに明らかになり一連の書物にまとめられた[36−40]．長い間被爆者という偏見のおそれから堅く閉ざしてきた口を若い高校生にひらいた漁船員のなまなましい証言から漁船員の被爆の惨状が明らかになっていった．漁師として日本人で初めて渡米したジョン・万次郎の土佐清水や中岡慎太郎の像が太平洋をのぞむ室戸は遠洋漁業が盛んで重要産業であった．室戸・土佐清水など高知県内のビキニ水爆被爆による遠洋漁船員と家族のくるしみは60余年後のいまも続き終わっていない[2]．第五福竜丸の冷凍

---

[2]筆者は2020年1月コロナ・パンデミックが始まる前に宿毛市を訪れ小野梓記念公園や吉田茂邸跡に近い宿毛文教センターで開かれていた子どもたちに高知県内漁船のビキニ水爆被爆の実相を伝えようと現職・退職教員でつくられた紙芝居『ビキニの海のねがい』の原画展(ビキニの海の紙芝居を作る会主催)を見た．証言をもとに生々し水爆と漁船員の実相が鮮やかに描かれている．これは2019年室戸市，2020年黒潮町，土佐清水市，高知市でも開かれた．筆者の名の一字は後に宰相となるこの吉田茂由来と両親から聞き，独立・繁栄の希望・「松の茂れる」香が籠る．湯川の次男・高秋は吉田茂にかかわる仕事に携わっていた．講談社で「この翻訳をもって吉田首相の真面目を伝え，その偉大な政治信条と国際平和に対する視野の広さを世界各国の人士に正しく理解してもらいたいと」企画された「一代の名宰相吉田茂氏の英語版評論的伝記の翻訳の仕事」に心臓麻痺で36歳で急逝する前夜まで携わっていたと，湯川スミ夫人の著書（[11]p.366）にある．

士だった大石又七さんが 2021 年 3 月亡くなった．1954 年に無線長だった久保山愛吉さんの被爆死から今も続く核の悲劇である．

湯川の「人生の転機」となったビキニ水爆に苦しむ高齢化した漁船員に関する記事が 68 年後の今も新聞に報じられている．2020 年夏，新聞に「ビキニ水爆実験巡る訴訟」と題する記事が載った[112]．

<u>「米国が 1954 年に太平洋・ビキニ環礁周辺で実施した水爆実験で，被曝（ひばく）した県内の元マグロ漁船員とその家族らが，・・・労災申請にあたる船員保険の適用を不認定とした処分の取り消し・・・を求めた訴訟の第 1 回口頭弁論が 31 日，高知地裁であった。（湯川うらら）」</u>

2021 年 1 月にも新聞記事が載った[113]．

<u>「核兵器禁止条約が発効した 22 日，1954 年に太平洋・ビキニ環礁周辺であった米国の水爆実験で被曝（ひばく）した県内の元マグロ漁船の関係者が，被災の実態調査を求める要望書を浜田省司知事に提出した。元漁船員を支援する「太平洋核被災支援センター」（宿毛市）の山下正寿事務局長が遺族らと県庁を訪れ，浜田知事に「高齢化が進み，ぎりぎりの状態。被災者のために力を貸してほしい」と訴え・・・県による被災の実態調査や資料収集などを求めた。　・・・一方で県による実態調査については『約 70 年が経ち，事実関係の調査は限界がある。国の責任で調査するべきで，要望する』と述べた。（湯川うらら）」．（下線は筆者）</u>

記事を執筆したのは夜須町の湯川胸像除幕式の小学生を探しあて除幕式の湯川秀樹先生のうつった写真（図 1，図 2，図 4）を私に見せてくれたあの若い朝日新聞社の女性記者・湯川うららである．稿を進めつつ二人の湯川の偶然の遭遇にわたしはなにか運命的なものを感じた．

理論物理学者・湯川秀樹と「原子力」．湯川の原子力に対する態度は1954年の毎日新聞声明で突然変わったように考えられ，その変化の経緯は明らかでなかった．それはその前の湯川の行動・考えが知られていなかったからである．高知県夜須町夜須小学校における湯川胸像の「再発見」，湯川の除幕式へ出席の写真の発見と高知訪問・講演内容の詳細の判明は湯川の平和思想の原点[111]を明らかにすることにつながるものであり，学術的にも意義は大きい．湯川の「生涯」転機」は高知訪問から始まっていた．平和の創造の湯川精神（湯川ガイスト）はのちの理論物理学者にも受け

継がれていく．原子核を知った理論物理学者が「原子力」にかかわる時代は超越的・宿命的なものがあった．湯川秀樹もその弟子の小林稔もまたその弟子の理論物理学者・永田忍(1928-2003)[114]もそうであり，筆者もこれら京都大学の先輩の諸先生からの影響のもとにあった[115]．湯川先生にあこがれて京都大学に進学し理論物理学の研究者の道に進んだ筆者には，湯川先生が筆者の故郷に胸像を残し，また平和主義の実践者として再度訪ねられていたことに偶然とはいいがたい運命を思い感慨こもるものがある．

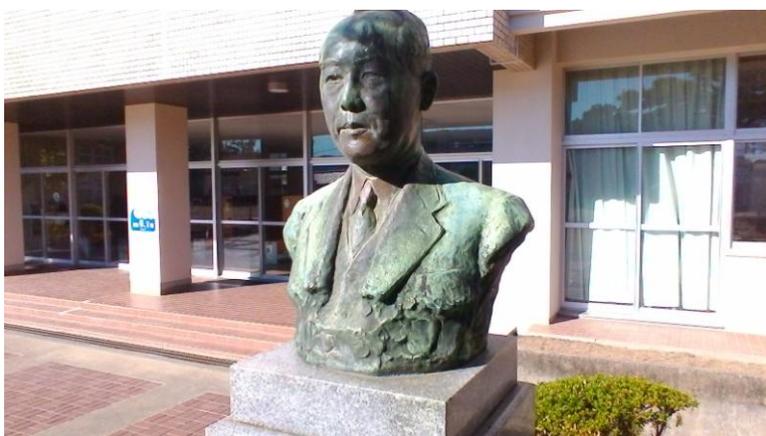

図27　高知県夜須小学校にある日本で最初の湯川秀樹先生胸像．（2018年 11 月 4 日筆者撮影）．

湯川胸像と平和の若葉の像，ふたつの像は湯川先生の生涯をあらわす双対かもしれない．湯川先生は偉大な物理学者として素粒子物理学をきりひらいただけでなく核のない世界連邦によるスケールのおおきい恒久平和を唱え志士のごとく20世紀をかけ抜けた．湯川先生がみずから眺めた高知の胸像（図 1）は六十有余年をへて21世紀になって世にひろく知られ（図27），いまその像じしんが双対性（自己双対性）を有し学問と平和の大切さを象徴しているように思える．その学問と人生について，寺田寅彦の弟子の渡辺慧はその追悼文で「みごとな人生」とかき，湯川の核廃絶運動を

ささえてきた素粒子物理学者・豊田利幸(1920-2009)は「学問と平和をつらぬくもの」を回想し，湯川の甥にあたる物理学者・小川岩男(1921-2006)は「アインシュタインの霊を追う如く」とのべている[116]．湯川先生は生涯，「胸像」については一切語らず，除幕式で「この胸像がみなさんの役にたてば大変光栄です」と述べたのみである．その言葉といまも夜須小学校校庭にたたずむ胸像には泉下の湯川先生の平和へのねがいが籠っていると思いを巡らせたい．

## 謝辞

## 参考文献

1. J. Chadwick, 「Possible existence of a neutron」, Nature 129, 312-312 (1932); J. Chadwick, 「The existence of a neutron, Proc. Royal Society, Series A, 136, 692-708 (1932).

2. W. Heisenberg, 「Über den bau der atomkerne I」, Zeit. Physik 77, 1-11 (1932); 「Über den bau der atomkerne II」, Zeit. Physik 78, 156-154 (1932); 「Über den bau der atomkerne III」, Zeit. Physik 80, 587-596 (1933).

3. D. Iwanenko, 「The neutron hypothesis」, Nature 129, 798-798 (1932).

4. H. Yukawa, 「On the interaction of elementary particles I」, Proc. Phys.-Math. Soc. Japan 17, 48-57 (1935).

5. C. M. G. Lattes, G. P. S. Occhialini, and C. F. Powell, 「Observations on the tracks of slow mesons in photographic emulsions.1」, Nature 160 (1947) 453-456; see also C. M. G. Lattes, H. Muirhead, G. P. S. Occhialini, and C. F. Powell, 「Processes involving charged mesons」, Nature 159, 694–697 (1947).

6. 「自然」増刊 追悼特集： 「湯川秀樹博士『人と学問』」p. 217（中央公論社 1981).

7. 大久保茂男 「湯川先生のはじめての胸像は何故高知に建てられたか」素粒子論研究・電子版 Vol. 28 (2019) No. 4 ; S. Ohkubo, arXiv: 1905.07707 「First bronze statue of Prof. Hideki Yukawa in Kochi」.

8. 大久保茂男 「1954 年高知・夜須小学校湯川胸像建立出席の湯川秀樹先生」徳島科学史雑誌 40, 47 (20121); S. Ohkubo 「Prof. Hideki Yukawa at the 1954 Unveiling Ceremony of the Yukawa Bronze Statue at School in Kochi」Journal of Tokushima Society for the History of Science 40, 47 (2021).

9. 湯川秀樹 『湯川秀樹著作集 別巻 対談 年譜・著作目録』（岩波書店 1990).

10. 夫人の名前については，先稿[7]で述べたように当時の新聞は本名の「澄子」としているので，本稿もそれに従う．後年「湯川スミ」として活躍するが，1954 年時点では「湯川スミ」存在しない．ふだんの生活でどう呼ばれていたかについては文献[11]に記載がある．京都知恩院にある「湯川家墓誌」には「平成

十八年五月十四日歿俗名澄子」とある.

11. 湯川スミ　『苦楽の園』(講談社 1976 年).

12. アブラハム・パイス(杉山滋郎・伊藤伸子訳)　『物理学者たちの 20 世紀　ボーア，アインシュタイン，オッペンハイマーの思い出』p. 481（朝日新聞社 2004).

13. 佐藤文隆　『ある物理学者の回想-湯川秀樹と長い戦後日本』(青土社　2019).

14. 大久保茂男　「湯川秀樹先生の初めての一般講演」素粒子論研究・電子版 Vol. 19 (2014) No. 3.（サイエンス・ポータル).

15. S. Ohkubo, J. Takahashi, and Y. Yamanaka, 「Supersolidity of the cluster structure in the nucleus $^{12}$C」, Prog. Theor. Exp. Phys. 041D012 (2020 April); J. Takahashi, Y. Yamanaka, and S. Ohkubo, 「Bose–Einstein condensation of dilute alpha clusters above the four-α threshold in $^{16}$O in the field theoretical superfluid cluster model」Prog. Theor. Exp. Phys. 093D03 (2020 Sept); S. Ohkubo, 「Existence of higher nodal band states with α+$^{48}$Ca cluster structure in $^{52}$Ti」 Phys. Rev. C 101, 041301(R) (2020).

16. 読売新聞　2020 年 1 月 9 日 夕刊, 2020 年 1 月 10 日 朝刊, 読売新聞　2020 年 2 月 4 日 朝刊.

17. 天野清　『量子力学史』(1948　日本科学社) p. 90.

18. S. Ohkubo,「Evidence of a higher nodal band α+$^{44}$Ca cluster state in fusion reactions and α clustering in $^{48}$Ti」Phys. Rev. C 104, 054310 (2021); Y. Hirabayashi and S. Ohkubo,「Existence of core excited $^8$Be$^*$ = α + α$^*$ cluster structure in α + α scattering」Prog.Theor. Exp. Phys. 113D02 (2021).

19. アメリカ帰りの湯川夫人の若々しいハイカラな印象・様子は，高知新聞 1954 年 3 月 24 日(水曜日)　朝刊「湯川夫人を囲む座談会」，高知新聞 1954 年 3 月 26 日(金曜日)　夕刊　「湯川澄子夫人の話」「日本婦人は社交性な　合理的な米国の都会生活」にも記されている.

20. 高知新聞 1954 年 3 月 23 日(火曜日)　夕刊 p. 3.

21. 除幕式で湯川秀樹の話を聞いた夜須小学校3年生のひとり(清達豊)は湯川に影響を受け京都大学理学部物理学科へ進む.　京都大学で湯川の講義「物理学通論」を受講，後に大阪で高等学校校長を務める.当時は全国的にも湯川にあこがれ京都大学に進む学生が多かった[13].　除幕を行った春樹(浜田)英子は大学(高知県立高知女子大学，現高知県立大学)へ進学.

22. 夜須町史編集委員会　『夜須町史』(上巻)（夜須町 1987)　p. 642.

23. 石川哲也「寺田寅彦の『X 線と結晶』から X 線自由電子レーザーへ」日本物理学会誌　70, 675 (2015).

24. 寺田寅彦(1909 年 5 月-1911 年 5 月に独・英へ留学 1916 年東京帝大教授)は 1912 年 6 月にラウエ(M. Von Laue 1879–1960)(1914 年ノーベル物理学賞)がゾンマーフェルト(A. J. Sommerfeld 1968-1951)のいるミュンヘン大学で 1912 年 X 線の回折現象を発見 [25] したと知ると，すぐに東京帝大医科大にあった X 線装置を借りて実験をおこない，1913 年 X 線の波動性による結晶による回折反射を発見. 論文は T. Terada, 「On the Transmission of X-rays through Crystals」Proceedings of the Tokyo Mathematico-Physical Society 2nd series. 7,60-70　(1913)で受理(5 月 3 日)掲載される. それ先立ってイギリスの科学論文誌 Nature に速報論文を送り，掲載されている. T. Terada, 「X-Rays and Crystals」Nature 91, 135–136 (1913)　(3 月 18 日投稿 4 月 10 日掲載), T. Terada, 「X-Rays and Crystals」Nature 91, 213 (1913)　(4 月 6 日投稿 5 月 1 日掲載). ノーベル賞級の研究であったが，1915 年ノーベル物理学賞は数か月早かったイギリスの Bragg 親子 (W. H. Bragg, W. L. Bragg ブラッグの法則を発見, W. H. Bragg「X-rays and Crystals」Nature 90, 572 (1913) (1 月 17 日投稿 1 月 23 日掲載，7 行の短報), W. H. Bragg「The Reflection of X-Rays by Crystals」Nature 91, 477 (1913) 6 月 10 日掲載 27 行の短報) におくられた. 寺田寅彦は本論文執筆後に W. L. Bragg から論文を受け取り，前年 1912 年 11 月 11 日に論文がケンブリッジ哲学学会 (Cambridge Philosophical Society) に受理され 1913 年 1 月 10 日出版されたことを知る， W. L. Bragg, 「The Diffraction of Short Electromagnetic Waves by a Crystal, Proceedings of the Cambridge Philosophical Society, 17, 43–57(1913). 寺田は「ラウエ映画の実験方法及其説明に関する研究」で 1917 年学士院恩賜賞受賞.

25. ミヒャエル・エッカルト　金子昌嗣　訳　『原子理論の社会史－ゾンマーフェルトとその学派を巡って』(海鳴社　2012)；Michael Eckert　『Die Atomphysiker: Eine Geschichte der theoretischen Physik am Beispiel der Sommerfeldschule』(Vieweg Verlag　1993).

26. 上田壽　『寺田寅彦断章』「寺田家の墓地のことなど」p. 201　(高知新聞社　1994).

27. 東京帝大で寺田寅彦の教えを受けた[26]菊池正士は 1928 年に大阪大で電子の波動性による回折を示した(菊池像). これもノーベル賞の研究であったが, 1 年ほど先んじ 1927 年に発見したイギリスの物理学者トムソン(G. P. Thomson　1892-1975) に 1937 年ノーベル物理学賞が授与された. 湯川は大阪大学で中間子理論を作り上げるとき実質的に菊池研究室に属しその自由な気風の研究での討論やコロキウムなどで薫陶を受けていた[28]ので, 湯川は寺田寅彦にとって孫弟子のような位置にある. 1949 年の湯川のノーベル賞受賞で, 寺田寅彦のＸ線回折研究での無念を果すものとなったが, 寺田は湯川が中間子論の論文が掲載された 1935 年大晦日に世を去っている. 湯川の先輩の京都大学の実験核物理学者, 荒勝文策は 1945 年無念にもアメリカ占領軍に自作のサイクロトロンを破壊され研究資料を没収されるが, 後年占領軍通訳に「後輩の湯川秀樹がノーベル賞を受賞した事で全てが埋め合わせさられた」と答えたとある(文献[54]の p. 246). 寺田が存命し湯川のノーベル物理学賞受賞を知ることができたならばいかに思ったことだろうか, 知る由もない.

28. 湯川秀樹　『湯川秀樹著作集　第7　回想・和歌　』p. 150　「菊池正士博士の追憶」(岩波書店 1989).

29. 湯川秀樹　『思想との対話　9　創造への飛躍』p. 9　「私の人生観の変遷」(講談社 1968).

30. 湯川が高知を訪問した理由は[7]で述べられている. 湯川は高知訪問にあたり 三高時代に校長であった高知県香南市・野市出身の物理学者・森総之助(1876(明治9)-1953(昭和28))について,「恩師故森総之助元三高校長に物理の手ほどきを受けたので高知は特になつかしい。」と語っている. 京都大学第 16 代総長の平沢興(1900(明治33)-1989(平成元年))は湯川に大きな影響を与えたのは森総之助であると語っている.「京都の三高に物理学の先生で、森総之助という素晴らしい独創的な物理学の先生がおられたんです。この三氏(筆者注、湯川秀樹、朝永振一郎、江崎玲於奈のこと)は、森先生の教え子です。つまり、ノーベル賞のもとは森総之助なんです。＜略＞この人は本当に学問好きな人。のちに、三高の校長になりましたが、これは自分はえらくない周囲から推されて、やむなくなった。校長としても偉い人だったが、まぁ、物理学の先生としては本当に学問を愛した人なんですね。」(平沢興『現代の覚者たち』(竹井出版 1988) p. 185-211). 湯川は平沢興と懇意であった. (平沢は湯川の長男春洋の仲人をしている。)湯川は森総之介については「三高には古くから、森総之助という有名な物理の先生がおられた。しかし、私が物理を習いはじめる時、この先生は外国へ行っていた。高校教授の外国出張は、珍しいことであった。」(『旅人 ある物理学者の回想』(朝日新聞社　1958)　p. 192[43])と記している. 森総之助は文部省留学生(1924 年5 月－1925 年 11 月)として 47 歳で留学, ミュンヘン大学ではＸ線を発見したレントゲン(W. Röntgen 1845-1923)の後任としていたヴィーンの法則を発見したヴィーン(W. Wien 1864-1928)のもとで研究する. 理論物理学者ゾンマーフェルトなど多くの学者と交流し、量子論やそれにつながる画期的な発見をおこなってきたドイツを中心とする欧米の進取の精神を学び湯川に伝えたのではなかろうか. 湯川は数学をならった高知出身の先生の印象も半生記『旅人』に記している.「数学の竹中馬吉先生・・・は、授業中に生徒を笑わせるのが非常に巧かった。土佐の人で、物理学校出だったそうだが＜略＞生徒は親愛感を抱いて、竹中先生を『馬さん』と呼んだ。」((『旅人　ある物理学者の回想』(朝日新聞社　1958)　p. 106)).

31. 湯川秀樹　『思想との対話　9　創造への飛躍』　p. 30「科学と人生論」(講談社 1968 年).

32. 高知新聞　1954 年 3 月 22 日(月曜日)　朝刊.

33. 高知新聞　1954 年 1 月 13 日(水曜日)　朝刊.

34. 高知新聞　1954 年 2 月 6 日(土曜日)　朝刊.

35. 高知新聞　1954 年 2 月 19 日(金曜日)　朝刊.

36. 高知県ビキニ水爆実験被災調査団 編　『もうひとつのビキニ事件 ： 1000 隻をこえる被災船を追う 』(平和文化 2004 年).

37. ビキニ幡多高校生ゼミナール/ 高知県ビキニ水爆実験被災調査団 編　『ビキニの海は忘れない―核実験被

災船を追う高校生たち』（平和文化　1988）.

38. 山下正寿　『核の海の証言　ビキニは終わらない』（新日本出版社　2012）.
39. 高知高校生ゼミナール編　『海光るとき』（民衆社　1990）.
40. ビキニ核被災ノート編集委員会,『ビキニ核被災ノート　隠された60年の真実を追う』（太平洋被災支援センター　2017）.
41. 朝日新聞 1954年3月2日　夕刊.
42. 朝日新聞 1954年3月14日　夕刊.
43. 朝日新聞 1954年3月16日　朝刊.
44. 読売新聞　1954年3月16日（火曜日）　朝刊.
45. 京都大学基礎物理学研究所湯川記念室ホームページ公開 湯川日記.
46. 高知新聞　1954年12月12日（日曜日）　朝刊.
47. 高知新聞　1954年3月23日（火曜日）　朝刊　p.3.
48. 高知新聞　1954年3月13日（土曜日）　朝刊　p.3.
49. 高知市中央公民館『高知市中央公民館26年史』（高知市中央公民館 1977）iii.
50. 高知市中央公民館『高知市中央公民館26年史』（高知市中央公民館 1977）　p.98 には湯川の演題は「科学者としての立場」（聴衆 2580名）と記録されている.
51. 高知新聞　1954年3月25日（木曜日）　夕刊　p.4.
52. 高知新聞　1954年3月23日（火曜日）朝刊　p.1（コラム小社会）.
53. 高知新聞　1954年3月24日（水曜日）　朝刊　p.4.
54. 政池明　『荒勝文策と原子核物理学の黎明』p.95-126（京大学術出版会　2018）.
55. 湯川秀樹「静かに思う」（週刊朝日　1945年11月）.
56. 湯川秀樹『自然と理性』「静かに思う」（秋田屋　1947年）.
57. 湯川秀樹『湯川秀樹著作集　巻5　平和への探求』p.3　「静かに思う」（岩波書店　1989年）.
58. 高知新聞 1954年3月6日（土曜日）　朝刊.
59. 山田一郎『坂本竜馬』p.10（新潮社1987）.
60. 高知新聞　1954年3月23日（火曜日）（p.1）の一面コラム記事「話題」を書いたのは松田忠吉で元共同通信科学部長の山田一郎の著書『坂本竜馬』（新潮社 1987）p.254 によると高知新聞社会部記者，高知新聞学芸部長を務めた.
61. 高知新聞 1954年3月23日（火曜日）　夕刊 p.1.
62. ラビー教授と湯川は交流があった．　湯川のアメリカ日記によると1939年9月16日（土曜日）ニューヨークで初めて会っている[63]．　9月18日には「別刷りを渡し、別刷りを貰う。＜略＞Rabi に Møller と Rosenfeld の letter（Natureにまで出ていない由）を見せてもらう。＜略＞一緒に中食に出かける。」とある．　湯川は1949年ラビーのコロビア大学に移り1953年帰国するまで勤める．コロンビア大学滞在時，湯川夫人もラビー夫人と交流があり，湯川が胃腸病で入院した時の話が[11]（p.273）にある.湯川の次男高秋はコロンビア大学で学ぶが「高秋はラビ夫妻などの深い情けを受けて、楽しく大学生活をつづけ」と夫人が記している（[11]p.316).　繰り込み群で1965年ノーベル物理学賞を受賞する理論物理学者 J. Schwinger（1918−1994）は教え子.
63. 湯川秀樹　『湯川秀樹著作集　巻7　回想・和歌　』p.171　（岩波書店　1989）.
64. ポアンカレ（吉田洋一訳）『科学と方法』　第1篇　学者と科学　第1章「事実の選択」　p.24（岩波文庫 1964）(Henri Poincaré) Sciences et Méthode (1908).
65. 寺田寅彦はポアンカレーの「事実の選択」と「偶然」を翻訳している．「事実の選択」は『東洋学芸雑誌』302巻　401号(1915年2月)p.13 に掲載．細谷暁夫『寺田寅彦『物理学序説』を読む』（窮理社　2020年）にも収録されている.
66. 寺田寅彦『寺田寅彦随筆集』（小宮豊隆編）岩波文庫　第1巻　「科学者と芸術家」　p.86-89(岩波書店 1993).

**67.** 大久保茂男『原子核のα粒子模型』パリティ 5 (4) p.48-50 (丸善 1990 年 4 月); 大久保茂男 「原子核の分子的構造：$^{4}$Ti 領域への展開」(日本物理学会誌 **45**(9) 648-650 1990); 大久保茂男「原子核の分子的構造：fp 殻領域への展開と武谷三段階論」 素粒子論研究 **76**(5) 137-146(1988); 大久保茂男 「アルファー・クラスターの夢を実証に 一山屋さんとの実験・理論の信頼の二人三脚一」 『山屋堯博士追悼文集 熱燗』p.3-18 (山屋堯博士追悼文集刊行会 1998 年 4 月); 大久保茂男 『原子核に副虹が存在』パリティ **30** (2) p.32-34 (丸善 2015 年 2 月); 科学新聞 2014 年 6 月 27 日「原子核に『2 番目の虹』存在」; 高知新聞 1989 年 8 月 15 日「重い核の陽子と中性子―ブドウの房状に存在」; 日本経済新聞 2007 年 5 月 12 日 in 「虹の「7 色」見えますか」; S. Ohkubo, Ed.「Alpha-clustering and molecular structure of medium-weight and heavy nuclei」 Progress of Theoretical Physics Supplement No. 132 (1998) ; 大久保茂男 「クラスター模型の展開」 青木健一・坂東昌子・登谷美穂子編「素粒子論研究」**112** 巻 6 号 (2006) 『学問の系譜―アインシュタインから湯川・朝永へ―』; 青木健一, 坂東昌子, 登谷美穂子 編, 『学問の系譜―アインシュタインから湯川・朝永へ―』(京都大学基礎物理学研究所発行 カラー版) pp.1-236 (2006); 大久保茂男 「独創性と美学」ふませすぷ 23 71-75 (2012).

**68.** 毎日新聞 1954 年 3 月 30 日 朝刊.

**69.** 湯川は 1939 年 9 月 21 にアメリカ・プリンストン高等研究所のアインシュタインに私邸ではじめて会う. 「Wheeler は Einstein の私宅に電話を掛けてくれ、すぐ会うことになり・・・Einstein の家に行く。大して立派でもない家の二階の書斎に通される。＜略＞風貌は写真で見た通りであるが大分老境に入ったように見える。Prof. Okaya の事や、日本へ来た時の思い出などを聞く。彼に相対性理論と量子力学との関係を聞くと彼は相変わらず後者が incompletely described picture に過ぎぬという。例えば dynamics から acceleration と言う概念を去り position と velocity だけを考えると、statistical law は得られない、などと語る。何だか遠い昔の世界へ引き戻されたような不思議な気持になる。しかしその眼には何ともいえない親しみ深みがある」[72]と記している。 湯川は 1948 年アインシュタインにいるプリンストン高等研究所に招待されて研究生活を送り、アインシュタインと親交を深める。 湯川は生涯の後半を広がりのある素粒子模型と非局所場理論の研究にささげ素領域理論に到達し、アインシュタインは後半生を電磁気力と重力の統一理論の研究にささげる。 ともに核廃絶と平和の創造に生涯をささげ二人の天才は共通するところがある。 「ラッセル―アインシュタイン宣言」(1955)に湯川は署名している。 湯川夫妻が 1948 年アメリカ・プリンストンにつくとアインシュタインが日本への原爆投下について涙を流し謝罪した話は湯川スミ夫人が著書[11] (p.199-200)に記している。『「アインシュタイン博士から私達二人に会いたいと電話がかかってきたから、午後僕の部屋に来なさい」との秀樹からの言葉に従って、研究所へ出かけた。窓から外をながめていると、広場のむこうから体格の立派な白髪の老人が、しっかいした足取りで、・・・やってきた。＜略＞アインシュタイン博士は私達の顔をみるなり、二人の手をご自分の両手で強く強くにぎりしめて、『罪もない日本人を原爆で殺戮して申しわけない』と涙をポロポロ流してわびられた。』. 湯川のアインシュタインとの交遊・思い出は『続天才の世界』(小学館 1973) p.7 にもある.

**70.** 朝日新聞 1954 年 4 月 16 日 夕刊. この談話で湯川は彼を 1948 年プリンストン高等研究所に招いた, かつて原爆開発のマンハッタン計画を物理学者として率いたオッペンハイマー(J. R .Oppenheimer 1904-1967 をふくめ誰ともと原子力(原爆)についてひとことも語ってないことをこの期に言明しているのは注目される。「オッペンハイマー博士が調べられているが、その背後になにがあるか、私は知らない。しかし私は博士を非常によく知っているので友人の一人として極めて嘆かわしいことに思う。私が米国にいたときには、全期間を通じて、原子力の応用についてだれかと話合ったということは、一つもない。私は理論物理学者である。応用については、ちっとも興味がない。」(筆者注、 オッペンハイマーは「原爆の父」といわれている。)

**71.** 新聞 ニューヨーク・タイムズ 1954 April 16.

**72.** 湯川秀樹 『湯川秀樹著作集 巻 7 回想・和歌 』p.178 (岩波書店 1989).

**73.** 湯川秀樹 『しばしの幸』(読売新聞社 1954 年 6 月 10 日発行).

**74.** 高知新聞 1954 年 3 月 24 日 (水曜日) 夕刊.

75. 高知新聞　1954 年 3 月 24 日（水曜日）　朝刊.
76. 高知新聞　1954 年 4 月 5 日（月曜日）夕刊.
77. 湯川秀樹　『湯川秀樹日記　昭和九年：中間子論への道』（朝日新聞社　2007）.
78. 湯川秀樹　『湯川秀樹日記 1945 京都で記した戦中戦後』 p.92（京都新聞出版センター 2020）.
79. 湯川秀樹　『湯川秀樹著作集　巻 7 回想・和歌 』p.286　「歌集 深山木」（岩波書店　1989）.
80. 湯川は帰국後,『量子力学序説』(1947 年 2 月初版, 弘文堂書房）の改訂増補版の『量子力学序説』(1954 年 6 月発行)の序文を 4 月に記している.
81. 湯川秀樹　『湯川秀樹著作集　巻 5　平和への探求』p.49（岩波書店　1989 年）.
82. 武谷三男　岩波新書 『死の灰』（岩波書店　1954）；『武谷三男著集 2　原子力と科学者』（勁草書房 1968）；　武谷三男　『原子力　武谷三男現代論集　1』（勁草書房　1974）.
83. 湯川秀樹　『続々天才の世界』p.8.　「荘子」（小学館 1979）.
84. 湯川秀樹　『湯川秀樹著作集　巻 5　平和への探求』（岩波書店　1989 年）.
85. 日本物理学会　第 8 回年会（1953 年 10 月 15 日－20 日　東大教養学部),基礎理論セッション(10 月 17 日午前）（日本物理学会誌第 8 号　p.443.
86. 日本物理学会誌　第 9 号　6 号 p.399.
87. 『京都大学基礎物理学研究所　1953~1978』（京都大学基礎物理学研究所　1978）.
88. 8 番目の内山龍雄(1916-1990)の講演は非可換ゲージ場(Non Abelian)理論の研究と思われる. 内山はこの発表の 2 週間後の 1 月末に 1954 年 9 月から, 湯川が 1948 年滞在したアメリカのプリンストン高等研究所に, 9 月からくるようにとの招待の手紙を受け取る[90]. 内山は一般ゲージ理論を 3 月までに完成させたが, 論文として投稿しなかった. 9 月にプリンストン高等研究所についた内山は, 大阪大学での卒業論文(湯川粒子の電磁相互作用の研究)の指導者であり[90], 1953 年からプリンストンに滞在中の小林稔先生(京都大学教授)に会い, ゲージ理論の話をすると, 同じことをヤン(C. N. Yang 1922- ）とミルズ（R. L. Mills 1927-1999)）がやっていることを聞く. その数日後, 論文のプレプリントを受け取り, 軽く目をとおすと自分の式と似ていることを知り愕然. 大きな衝撃を受け詳しく見ることなく半年間放置する. 翌 1955 年 1 月になり滞在が一年延長されることになり, それまでやめていたゲージ場の研究に戻る決意をし, ヤン・ミルズの論文を詳しく読むと, 自分のゲージ理論がヤン・ミルズ理論にはない重力場をふくむはるかに進んだ一般的なゲージ理論であること知る. 内山が気を取り直しプリンストン高等研究所から論文を投稿するのは 1955 年 7 月 7 日である. 内山の重力を含んだ画期的なゲージ場理論が論文として公表されるのは翌年 1956 年. (Ryoyu Utiyama,「Invariant Theoretical Interpretation of Interaction」Physical Review 101, 1597-1607). 内山は渡米前 1954 年 3 月に論文を投稿しなかったのが悔やまれると述べる. Yang による強い相互作用の非可換ゲージ場についての研究発表(ヤン・ミルズ理論)のセミナーは 1954 年 2 月 23 日火曜日に行われた. セミナーが始まるとゲージ粒子の質量が 0 になる問題があること知っていたパウリ(W. E. Pauli 1900-1958）がくりかえし質問し, ヤンは答えられず, セミナー会場には気まずい雰囲気が漂る. オッペンハイマー（J. R .Oppenheimer 1904-1967）がとりなして何とか終えた（[91] p.183.). 内山の 1954 年 1 月の基礎物理学研究所での発表はヤンのセミナー講演より早い. ヤン・ミルズの論文が米国学会誌フィジカル・レビュー誌に掲載されるのは 1954 年 6 月 28 日（C. N. Yang and R. L. Mills, 「Conservation of Isotopic Spin and Isotopic Gauge Invariance」Physical Review 96, 191-195) である. オラファッチは文献 [91] で内山のゲージ理論に 1 つの章をあて内山の方が早く非可換ゲージ場理論に到達していたが, 論文公表では Yan-Mills が早かったとしている. Yang は 1957 年ノーベル物理学賞受賞（素粒子におけるパリティ非保存発見). 内山は異国の地で 1913 年寺田寅彦が先行するブラッグの研究を知り衝撃を受けたると同じ衝撃を 41 年後に経験する. 寺田寅彦は論文最後にブラッグの先行性をはっきり認める脚注を加え, 痛恨ごとは何も語らず, X 線研究を大学院生の西川正治（1884-1952）に託した. 内山は後世のため, 痛恨記（[92] p.209-216　10 章　痛恨記）を残した. 研究者は多かれ少なかれ似た経験を味わい, 乗り越えていく宿命にある.

89. 高林武彦 「自然」増刊 追悼特集：「湯川秀樹博士『人と学問』」p. 38 （中央公論社 1981）.

90. 内山龍雄「迷想記 （統一場理論に誘われて）」素粒子論研究 **82**, No.6 p. 494 （1990）.

91. L. O'Raifeartaigh, 『The Dawn of Gauge Theory』（Princeton University Press 1997）.

92. 内山龍雄 『物理学はどこまで進んだか―相対論からゲージ論へ』（岩波書店 1987）.

93. 湯川秀樹 『湯川秀樹著作集 巻9 学術篇 II』 p. 261 （岩波書店 1989）の第9章「素粒子と時空」で素領域の定式化があたえられており、§9.1は「ひろがりをもつ素粒子と素領域」である。 湯川自身による素領域概念の説明は、湯川秀樹、『湯川秀樹著作集 巻3 物質と時空』「『素領域理論』とは何か」p. 44 （岩波書店 1989）にある。 原著論文は、Hideki Yukawa「Space-Time Description of Elementary Particles」*Proceedings of the International Conference on Elementary Particles*, Kyoto, 24-30 September 1965 （Progress of Theoretical Physics, 139-158 （1966））. 湯川の非局所場理論, 素領域理論を引用した最近の研究は文献[97]に見られる.

94. Hideki Yukawa, 「Atomistics and the Divisibility of Space and Time」[Progress of Theoretical Physics 37/38 512-523.](#)

95. Yasuhisa Katayama and Hideki Yukawa, 「Field Theory of Elementary Domains and Particles I」[Progress of Theoretical Physics, Supplement 41, 1-21 (1968)](#) .

96. Yasuhisa Katayama, Isao Umemura and Hideki Yukawa, 「Field Theory of Elementary Domains and Particles II」[Progress of Theoretical Physics, Supplement 41, 22-55 (1968)](#) .

97. K. Aouda, S. Nakada and H. Toyoda, 「An Approach to Yukawa's Elementary Domain Based on AdS5 Spacetime」ArXiv:1710.09677v5 (hep-ph) （2018）. K. Aouda, S. Nakada and H. Toyoda, 「Bi-local fields in AdS5 spacetime.」Journal of High Energy Physics **90** (2016).

98. 湯川の戦後の非局所場理論, 素領域理論の研究について理論物理学者がド・ブロイ（Louis de Broglie 1892-1987）、ハイゼンベルグに学び, 湯川が東大に移る計画では助教授として予定され, 湯川と戦前から交流の長かった渡辺慧(1910-1993)は湯川追悼に寄せた文章で次のように述べて評価している（『自然』増刊 追悼特集 湯川秀樹博士『人と学問』p. 46 「〈湯川さんの非局所場は、明確な "哲学" をもっており、数学的にも独創的であり、その真実性は、実験データに直接合わなくても、人を打つものを持っています。今は認められてなくても、やがて別の衣を着せられて甦って来るでしょう。これは中間子論以上の大事業です。こういう独創的な仕事はノーベル賞にはなりません。」 渡辺は湯川の学問に対する態度についても書いている。 「私は、昔から学者のタイプを二つに分けて、一方をN型（長岡のN）、と他方をT型（寺田のT）にして見ますと、その行動の構造がよく解るような気がしています。N型は研究の動力を成功欲、権力欲から導いてくる型であり、T型はそれを、好奇心、学問自身の楽しみから導いて来る型です。アメリカの学者は、ホィーラー（J. Wheeler）さんなどを除けば大体N型が多いようです。日本では、T型がまだ割り合い多く見受けられます。その成功や高名度にもかかわらず、湯川さんとか・・・は確かにT型です。朝永さんはその落語的飄逸さの見かけにも関わらず明らかにN型です。」（長岡は東京帝大教授の長岡半太郎(1865-1950)（1893年5月-1896年9月, 独へ留学）で原子の有核土星型模型（H. Nagaoka, 「Kinetics of a system of particles illustrating the line and the band spectrum and the phenomena of radioactivity」, [Phil. Mag. Ser. Series 6, Vol. 7, 445-455 (1904)](#); H.Nagaoka「On a dynamical system illustrating the spectrum lines and the phenomena of radio-activity 」, [Nature 69, 392-393 (1904)](#) を提唱した. T型の寺田は寺田寅彦のこと、東京帝大教授寺田寅彦（1878-1935）でX線の結晶による回折反射発見.

99. 共同通信元科学部長・京都支局長で湯川秀樹を担当した山田一郎(1919-2010)は著書『坂本竜馬』（新潮社 1987）の最終頁 p. 263 で湯川のメディアにおける大きさについて次のように述べて本を締めくくっている。 「京都北白川、湯川記念館の基礎物理学研究所に、私は(後任の)今井を連れて行った。湯川さんに離任着任のあいさつをするためであった。湯川さんの存在は当時、ジャーナリズムの世界でもそれほど大きかったのである。『湯川さんを確かに引き継いだよ』近くの喫茶店駿々堂で私は今井にそう言った。」山田一郎が京都支局長を離任するのは1963年4月. 後任者今井幸彦は共同通信で科学部長をつとめ、京都支局長時代は湯川を担当. 湯川の死去の1週間後1981年9月14日没. 坂本竜馬の暗殺者今井信郎の孫と本書にある. 山田一

郎は高知市仁井田出身で筆者とは同郷である．寺田寅彦の研究でも知られ，寺田寅彦のルーツを明らかにしたことはよく知られている．ルーツの集落には石碑が建てられている．1954 年 3 月桂浜で坂本竜馬を見上げた湯川がその暗殺者の末裔とされる人物に 9 年後京都の基礎物理学研究所の所長室で会うというのも奇縁である．湯川番をつとめた今井幸彦元京都支局長・科学部長には『坂本竜馬を斬った男 幕臣今井信郎の生涯』（新人物往来社）の著書がある．

100. 高知新聞 1969 年 6 月 22 日（日曜日） 朝刊.
101. 高知新聞 1969 年 6 月 11 日（水曜日） 朝刊.
102. 高知新聞 1969 年 6 月 22 日（日曜日） 朝刊.
103. 紀貫之 『土佐日記』（岩波書店 1979）；『続日本文学大系 24』「土佐日記 蜻蛉日記 紫式部日記 更科日記」（岩波書店 1996）.
104. 湯川秀樹 『湯川秀樹著作集 巻 6 読書と思索』p. 23「荘子」（岩波書店 1989）.
105. 湯川秀樹 『天才の世界』（小学館 1973）.
106. 高知新聞 1969 年 6 月 20 日（金曜日） 夕刊.
107. 高知新聞 1969 年 6 月 23 日（月曜日） 朝刊.
108. 筆者が湯川の講義で聞いた観測理論は『湯川秀樹著作集 巻 3 物質と時空』（p. 187. ）「観測の理論」にある.
109. 高知市中央公民館『高知市中央公民館 26 年史』（高知市中央公民館 1977）p. 91.
110. 小林稔の講演概要は高知新聞 1964 年 8 月 4 日（火）「夏期大学ノート 素粒子の世界」にある．湯川の弟子で中間子理論の建設に協力した坂田昌一（名古屋大学教授）もこの高知市の夏期大学（第 8 回）で小林稔に先立って 1958 年 8 月 2 日「宇宙時代と原子力」と題して講演している(高知市中央公民館『高知市中央公民館 26 年史』（高知市中央公民館 1977）p. 88).
111. 田中正『湯川秀樹とアインシュタイン 戦争と科学の世紀を生きた科学者の平和思想』（岩波書店 2008）.
112. 朝日新聞 2020 年 8 月 1 日 朝刊.
113. 朝日新聞 2021 年 1 月 23 日 朝刊.
114. 「原子核研究」**50**, No. 5 （2006）.
115. 大久保茂男「原子核の分子的構造と核間相互作用 ―永田忍さんの教えをうけて―」「原子核研究」**50**, No. 5, 35 （2006）.
116. 「自然」増刊 追悼特集：「湯川秀樹博士『人と学問』」p. 217（中央公論社 1981）.

## 図・写真の出典

図 1―図 7：1954 年夜須小学校配布物，図 8：清藤禮次郎 2021 年写真撮影，図 9：夜須小学校公開展示物（清藤禮次郎 2021 年写真撮影），図 10：清藤禮次郎写真撮影，図 11―図 12：夜須町公民館公開展示・配布物，図 13：1981 年夜須小学校配布物，図 14：大久保茂男 2018 年撮影，図 15―図 18：説明文記載引用文献，図 19：大久保茂男 2018 年撮影，図 20―図 26：説明文記載引用文献，図 27：大久保茂男 2018 年撮影．

[i] ohkubo@rcnp.osaka-u.ac.jp Research fellow of RCNP （高知県立大学名誉教授, Emeritus Professor of The University of Kochi）